\documentclass[journal,onecolumn]{IEEEtran}
\ifCLASSINFOpdf
\else
\fi

\usepackage[english]{babel}
\usepackage[T1]{fontenc}
\usepackage{amsmath,amssymb,amsthm,booktabs,enumitem,array,tabularx}
\usepackage[numbers,sort&compress]{natbib}
\usepackage{needspace}
\usepackage{microtype}
\usepackage{xcolor}
\definecolor{linkblue}{RGB}{0,0,145}
\usepackage[colorlinks=true,linkcolor=linkblue,citecolor=linkblue,urlcolor=linkblue]{hyperref}
\hypersetup{pdftitle={Catalytic compression of classical communication in exact entanglement transformations},pdfauthor={Vito Viesti and Alexander Streltsov}}
\setlist{itemsep=2pt,topsep=3pt,leftmargin=*}
\allowdisplaybreaks[1]
\newtheorem{theorem}{Theorem}[section]
\newtheorem{lemma}[theorem]{Lemma}
\newtheorem{proposition}[theorem]{Proposition}

\theoremstyle{definition}

\theoremstyle{remark}

\newcommand{\Bto}{B_{\rightarrow}}
\newcommand{\Cst}{C^{*}}

\newcommand{\spec}{\operatorname{spec}}

\begin{document}

\title{Catalytic compression of classical communication
in exact entanglement transformations}

\author{Vito Viesti and Alexander Streltsov
\thanks{This work was supported by the National Science Centre,
Poland, under Grants 2022/46/E/ST2/00115
and 2024/55/B/ST2/01590.}
\thanks{Vito Viesti and Alexander Streltsov are with
the Institute of Fundamental Technological Research,
Polish Academy of Sciences, Pawi\'nskiego 5B,
02-106 Warsaw, Poland
(e-mail: \href{mailto:vviesti@ippt.pan.pl}{vviesti@ippt.pan.pl};
\href{mailto:astrel@ippt.pan.pl}{astrel@ippt.pan.pl}).}}

\maketitle

\begin{abstract}
Entanglement catalysis is usually associated with making otherwise
impossible transformations possible. Here we show that it can also
change how much classical communication a transformation requires.
A catalyst, returned exactly and uncorrelated with the output, can
reduce the message alphabet size required for exact transformations
of bipartite pure states by a factor that grows without bound
as the Schmidt rank increases. In the asymptotic many-copy regime, however, this advantage disappears at the level of the
leading communication rate: the optimal catalytic and noncatalytic
rates coincide. We characterize the optimal catalytic cost in terms of
Rényi entropies, revealing how the entanglement structure of the source
and target governs the finite-copy advantage. Our results establish
reusable entanglement as a resource for compressing classical
communication in quantum state manipulation.
\end{abstract}


\IEEEpeerreviewmaketitle

\section{Introduction}

Manipulating entanglement between distant laboratories requires local
control and, in general, the exchange of classical messages. For
bipartite pure states, Nielsen's theorem gives a complete criterion for
whether a deterministic transformation is possible under local operations
and classical communication (LOCC): the source Schmidt vector must be
majorized by the target Schmidt vector~\cite{Nielsen1999}. The theorem
leaves open how much communication is needed to implement the
transformation. Different protocols for the same conversion can require
different message alphabet sizes, making classical communication a resource
to be optimized in its own right
\cite{LoPopescu2001,HaydenWinter2003,HarrowLo2004}.

Entanglement catalysis adds a further possibility to this optimization.
The parties may use an auxiliary entangled state, provided that it is
recovered at the end of the protocol. Catalysis was introduced through
the observation that such assistance can enable transformations that
would otherwise be impossible~\cite{JonathanPlenio1999}. The resulting
convertibility relation has an entropic characterization
\cite{Klimesh2007,Turgut2007}, and finite sufficient tests have also been
developed~\cite{ElkoussMaityNemaStrelchuk2026}.
Broader accounts of catalytic resources and return conditions
can be found in
Refs.~\cite{DattaKondraMillerStreltsov2023,LipkaBartosikWilmingNg2024}. Here we investigate what this
auxiliary resource can do for a transformation that is already possible.
Even when no catalyst is needed for feasibility, access to one may allow
the parties to coordinate their local operations using a smaller message alphabet.

To isolate this effect, we consider exact deterministic
transformations between bipartite pure states with distinct,
full-support Schmidt vectors $p,q$ of equal length.
We assume that the conversion is already possible by LOCC
without a catalyst. Every catalyst is a finite-dimensional pure
bipartite state returned exactly and uncorrelated with the target.
These return conditions matter: approximate target preparation
with exact catalyst-marginal return can lead to a convertibility
criterion governed by entanglement entropy
\cite{KondraDattaStreltsov2021}, while allowing approximate
catalyst return permits entanglement embezzlement
\cite{vanDamHayden2003}. Our requirement of an exact product output ensures that the
auxiliary entanglement remains fully available for reuse
after each conversion.

We quantify classical communication by the alphabet
size. In a one-way LOCC protocol, the alphabet consists
of the possible classical labels sent by Alice
to Bob. For a finite-round protocol, it consists
of the complete sequences of exchanged classical
messages that occur with nonzero probability. We allow local ancillas and uncommunicated local
outcomes. Under these assumptions, the optimum
can be attained with one-way communication
(Appendix~\ref{app:oneway}).
We denote the minimum one-way alphabet size by
$\Bto(p\to q)$ and its minimum over finite pure
catalysts by $\Cst(p\to q)$.
This perspective complements studies of pure-state
conversion without communication under local
operations with shared randomness
\cite{GeorgeChitambar2024} and catalytic local
operations with bounded catalyst Schmidt number
and marginal return~\cite{LipkaBartosik2026}.

The freedom to choose a catalyst turns the communication optimization
into an entropic threshold. We prove that
\begin{align}
    \Cst(p\to q)
    =\left\lfloor2^{\Delta_R(p,q)}\right\rfloor+1,
    \label{eq:intro-main-result}
\end{align}
where
\begin{align}
    \Delta_R(p,q):=
    \max_{\alpha\in[0,\infty]}
    [H_\alpha(p)-H_\alpha(q)]
\end{align}
is the largest R\'enyi entropy difference between source and target.
Logarithms are in base two, with the endpoint entropies
defined by continuity. Equivalently, an exact catalytic
protocol with message alphabet size at most $m$
exists if and only if
\begin{align}
    m>2^{\Delta_R(p,q)}.
\end{align}
Thus the optimal alphabet size is the smallest integer
strictly larger than $2^{\Delta_R(p,q)}$.
In particular, when this quantity is itself an integer,
an alphabet of that size is insufficient.
Every admissible budget $m$ can be achieved with
a finite-dimensional catalyst.

The size of the resulting advantage can grow without bound
as $d$ increases. For every $d\geq3$, we construct a transformation whose
minimum unassisted message alphabet size is $d$, the largest
possible for Schmidt vectors of length $d$, whereas its
minimum catalytic message alphabet size is two.

This large saving for individual transformations raises the question
of how the advantage behaves when many copies of a fixed pair are
processed together. We show that collective processing achieves the
same asymptotic communication rate with or without a catalyst, while
retaining exactness. Catalysis can therefore reduce the communication
needed for individual conversions without changing the leading cost
per copy in the many-copy limit.

Proving the catalytic characterization requires turning entropy inequalities into
an exact protocol with a controlled message alphabet. The unitary-orbit
characterization of communication cost and the equalization of
decomposition weights due to Daftuar and
Hayden~\cite{DaftuarHayden2005} translate the problem into one of spectral
allocation. We construct suitable allocations for many copies and
realize them in finite dimensions while preserving the source spectrum
exactly. To recover a single-copy protocol, we build on the connection
between multicopy transformations and catalysis established by Duan
et al.~\cite{DuanFengLiYing2005}. Our construction adds
control of communication: an $m^k$-term catalytic decomposition on $k$
copies yields an $m$-term decomposition on one copy. This is the step
that converts a many-copy entropic criterion into an attainable
single-copy communication budget.

We develop the operational framework and state the catalytic characterization
in Sec.~\ref{sec:framework}, before illustrating the communication saving
in Sec.~\ref{sec:maximal-compression}. The proof of the catalytic
threshold occupies Secs.~\ref{sec:proof-characterization}
and~\ref{sec:allocation}; its asymptotic consequences are developed in
Sec.~\ref{sec:rates}. 

\section{Operational framework and catalytic cost}
\label{sec:framework}

\subsection{Communication cost and unitary-orbit characterization}
\label{subsec:communication-cost}

We consider exact deterministic transformations between
bipartite pure states by local operations and classical
communication (LOCC). Up to local unitaries, the source
and target states can be written as
\begin{align}
    |\psi_p\rangle
    &= \sum_{i=1}^{d}\sqrt{p_i}\,|ii\rangle,
    \\
    |\phi_q\rangle
    &= \sum_{i=1}^{d}\sqrt{q_i}\,|ii\rangle,
\end{align}
where $p,q\in\mathbb{R}^d$ are probability vectors
with strictly positive entries, arranged in
nonincreasing order.

Throughout, we assume $p\neq q$ and $p\prec q$,
where majorization means
\begin{align}
    \sum_{i=1}^{\ell}p_i
    \leq
    \sum_{i=1}^{\ell}q_i,
    \qquad 1\leq\ell<d.
    \label{eq:majorization}
\end{align}
By Nielsen's theorem~\cite{Nielsen1999}, this
condition is necessary and sufficient for the
transformation $|\psi_p\rangle\to|\phi_q\rangle$
to be achievable by deterministic LOCC.
Our interest lies in the amount of classical
communication required to implement it.

We quantify classical communication by the alphabet
size, defined as the number of distinct complete
sequences of exchanged classical messages occurring
with nonzero probability. We consider finite-round
LOCC protocols and allow arbitrary finite-dimensional
local ancillas and uncommunicated local outcomes.
In particular, different local outcomes may
correspond to the same communicated message.

For the pure-state transformations considered here,
restricting to one-way communication entails no loss
of generality under this communication convention.
Indeed, a finite-round LOCC protocol can be reduced
to a one-way protocol without increasing the
communication alphabet size. A proof of this
reduction, building on the results of Lo and
Popescu~\cite{LoPopescu2001} and Harrow and
Lo~\cite{HarrowLo2004}, is given in
Appendix~\ref{app:oneway}.

We may therefore assume that Alice performs a local
instrument and communicates a classical message $j$
to Bob, who applies a local operation conditioned
on the received message. The protocol must produce
the prescribed target state exactly for every
message occurring with nonzero probability.

For a protocol $\Pi$, let $\mathcal J_\Pi$ denote
the set of possible communicated labels, and let
$N(\Pi):=|\mathcal J_\Pi|$ be its cardinality.
To identify the label, Alice needs a binary encoding
that distinguishes among these $N(\Pi)$ possibilities.
Since $L$ bits encode at most $2^L$ distinct values,
the minimum fixed-length encoding requires
$\lceil\log_2 N(\Pi)\rceil$ bits.

We define the minimum message alphabet size as
\begin{align}
    B_{\rightarrow}(p\to q)
    :=\min_{\Pi}N(\Pi),
    \label{eq:communication-cost}
\end{align}
where the minimum runs over all exact deterministic
one-way LOCC protocols implementing the transformation.

The corresponding logarithmic communication cost is
$\log_2 B_{\rightarrow}(p\to q)$, while the minimum
number of transmitted bits with fixed-length binary
encoding is $\lceil\log_2 B_{\rightarrow}(p\to q)\rceil$.

The communication cost admits a characterization
in terms of convex combinations of unitary
conjugates of the target reduced state, following
the framework of Daftuar and
Hayden~\cite{DaftuarHayden2005}.
Let
\begin{align}
    P=\operatorname{diag}(p),
    \qquad
    Q=\operatorname{diag}(q).
\end{align}
For every integer $m\geq 1$, we have
\begin{align}
    B_{\rightarrow}(p\to q)\leq m
    \quad\Longleftrightarrow\quad
    P=\sum_{j=1}^{m}w_jU_jQU_j^\dagger,
    \label{eq:unitary-characterization}
\end{align}
for some unitary matrices $U_j\in\mathrm{U}(d)$
and nonnegative weights $w_j$ satisfying
$\sum_jw_j=1$.

This equivalence holds for general local
instruments, including ancillary systems and
uncommunicated outcomes. For completeness,
we provide a proof under our operational
convention in Appendix~\ref{app:operational}.

Furthermore, by Corollary 2.7 of
Ref.~\cite{DaftuarHayden2005}, any such
decomposition can be replaced by one with
equal weights and the same number of terms.
Consequently,
\begin{align}
    B_{\rightarrow}(p\to q)\leq m
    \quad\Longleftrightarrow\quad
    P=\frac{1}{m}\sum_{j=1}^{m}U_jQU_j^\dagger.
    \label{eq:uniform-characterization}
\end{align}

For a finite-dimensional Hermitian matrix $X$,
we define
\begin{align}
    \mathcal{F}_m(X)
    :=
    \left\{
    \frac{1}{m}\sum_{j=1}^{m}U_jXU_j^\dagger:
    U_j\ \text{unitary}
    \right\}.
    \label{eq:unitary-average-set}
\end{align}
The noncatalytic communication cost can therefore
be expressed as
\begin{align}
    B_{\rightarrow}(p\to q)
    =
    \min\left\{
    m\geq 1:
    P\in\mathcal{F}_m(Q)
    \right\}.
    \label{eq:communication-optimization}
\end{align}
Thus, minimizing the classical communication
required for an exact pure-state transformation
is equivalent to finding the smallest number of
unitary conjugates of the target reduced state
whose uniform average reproduces the source
reduced state.

\subsection{Elementary properties of the
one-way communication cost}
\label{subsec:cost-properties}

We collect a few properties of
$B_{\to}$ that will be used in
the constructions below.

\emph{Dimension bound.}
For probability vectors $x,y$
of length $D$, with $x\prec y$,
\begin{align}
    \Bto(x\to y)\leq D.
    \label{eq:majorization-message-bound}
\end{align}
Indeed, by majorization and
Carath\'eodory's theorem, $x$ is
a convex combination of at most
$D$ permutations of $y$.

\emph{Composition and unchanged factors.}
Sequential transformations satisfy
\begin{align}
    \Bto(x\to z)
    \leq
    \Bto(x\to y)\Bto(y\to z).
    \label{eq:composition-cost}
\end{align}
Indeed, suppose that
\begin{align}
    X=\sum_{j=1}^{m}a_jU_jYU_j^\dagger,
    \qquad
    Y=\sum_{k=1}^{\ell}b_kV_kZV_k^\dagger.
\end{align}
Substitution gives
\begin{align}
    X=\sum_{j=1}^{m}\sum_{k=1}^{\ell}
    a_jb_k(U_jV_k)Z(U_jV_k)^\dagger,
\end{align}
which is a convex unitary decomposition
with at most $m\ell$ terms.

Similarly, an unchanged tensor factor
does not increase the communication cost:
\begin{align}
    \Bto(c\otimes x\to c\otimes y)
    \leq\Bto(x\to y).
    \label{eq:unchanged-factor}
\end{align}
Indeed, any decomposition
\begin{align}
    X=\sum_{j=1}^{m}a_jU_jYU_j^\dagger
\end{align}
immediately yields
\begin{align}
    C\otimes X
    =
    \sum_{j=1}^{m}a_j
    (I\otimes U_j)
    (C\otimes Y)
    (I\otimes U_j^\dagger),
\end{align}
where $C=\operatorname{diag}(c)$.
Thus the same number of unitary
terms suffices.

\emph{Direct-sum rule.}
Let $x_i,y_i$ be pairs of
probability vectors of equal length,
and let $\lambda_i>0$ satisfy
$\sum_i\lambda_i=1$.
Then
\begin{align}
\begin{aligned}
    &\Bto\left(
    \bigoplus_i\lambda_i x_i
    \to
    \bigoplus_i\lambda_i y_i
    \right)
    \leq
    \max_i\Bto(x_i\to y_i).
\end{aligned}
    \label{eq:direct-sum-budget}
\end{align}

To prove this, set
$M=\max_i\Bto(x_i\to y_i)$
and write
$X_i=\operatorname{diag}(x_i)$,
$Y_i=\operatorname{diag}(y_i)$.
By the operational characterization
and \cite[Corollary~2.7]{DaftuarHayden2005},
each block admits an equal-weight
decomposition
\begin{align}
    X_i=\frac1M\sum_{\ell=1}^{M}
    U_{i,\ell}Y_iU_{i,\ell}^{\dagger}.
\end{align}
Defining
$U_\ell=\bigoplus_i U_{i,\ell}$,
we obtain
\begin{align}
\begin{aligned}
    \bigoplus_i\lambda_iX_i
    =\frac1M\sum_{\ell=1}^{M}
    U_\ell
    \left(\bigoplus_i\lambda_iY_i\right)
    U_\ell^\dagger.
\end{aligned}
\end{align}
Thus all blocks share the same
$M$ message labels.

\subsection{Catalytic communication cost}
\label{subsec:catalytic-cost}

We now investigate whether entanglement catalysis
can reduce the classical communication required
for a transformation that is already achievable
without catalytic assistance.

To this end, we allow Alice and Bob to share an
additional bipartite pure state
\begin{align}
    |\chi_r\rangle
    =
    \sum_{a=1}^{h}\sqrt{r_a}\,|aa\rangle,
\end{align}
where $r=(r_1,\ldots,r_h)$ is a strictly positive
probability vector of arbitrary finite length $h$.
The state $|\chi_r\rangle$ acts as a catalyst if
the transformation
\begin{align}
    |\psi_p\rangle\otimes|\chi_r\rangle
    \longrightarrow
    |\phi_q\rangle\otimes|\chi_r\rangle
    \label{eq:catalytic-transformation}
\end{align}
can be implemented exactly and deterministically.
In particular, the catalyst must be returned
unchanged and uncorrelated with the target state.

For a fixed catalyst $r$, the minimum message alphabet
size is $B_{\rightarrow}(p\otimes r\to q\otimes r)$.
Optimizing over catalysts, we define the minimum
catalytic message alphabet size as
\begin{align}
    C^*(p\to q)
    :=
    \min_r B_{\rightarrow}
    (p\otimes r\to q\otimes r),
    \label{eq:catalytic-cost}
\end{align}
where the minimum is taken over all strictly
positive probability vectors of arbitrary finite
length. Since the trivial catalyst $r=(1)$ is
included, we have
\begin{align}
    C^*(p\to q)\leq B_{\rightarrow}(p\to q).
\end{align}

The unitary-orbit characterization established
in Sec.~\ref{subsec:communication-cost} provides
an equivalent formulation of this optimization.
Writing $R=\operatorname{diag}(r)$, we obtain
\begin{align}
    \begin{aligned}
        C^*(p\to q)
        &=\min_R\min\bigl\{m\geq1:\, P\otimes R\in\mathcal F_m(Q\otimes R)\bigr\},
    \end{aligned}
    \label{eq:catalytic-optimization}
\end{align}
where $R$ ranges over all positive definite
density matrices of arbitrary finite dimension.

We show that this optimization is completely
characterized by the Rényi entropy differences
between the source and target Schmidt vectors.
For a probability vector $x$, the Rényi entropy
of order $\alpha$ is defined by
\begin{align}
    H_\alpha(x)
    =
    \frac{1}{1-\alpha}
    \log_2\left(\sum_i x_i^\alpha\right),
    \qquad \alpha>0,\quad\alpha\neq 1,
\end{align}
with the continuous extensions
\begin{align}
    H_0(x)&=\log_2|\operatorname{supp}x|,\\
    H_1(x)&=-\sum_i x_i\log_2x_i,\\
    H_\infty(x)&=-\log_2\max_i x_i.
\end{align}

We define the maximum R\'enyi entropy difference by
\begin{align}
    \Delta_R(p,q)
    :=\max_{\alpha\in[0,\infty]}
    \bigl[H_\alpha(p)-H_\alpha(q)\bigr].
    \label{eq:renyi-gap}
\end{align}
Since $p$ and $q$ are arranged in nonincreasing
order, $p\neq q$ implies that they are not
permutations of one another. Together with
$p\prec q$, strict Schur concavity of the
Shannon entropy therefore gives
\begin{align}
    \Delta_R(p,q)
    \geq H_1(p)-H_1(q)>0.
\end{align}

Our main result establishes the exact optimal
catalytic communication cost.

\begin{theorem}[Exact catalytic communication cost]
\label{thm:catalytic-cost}
Let $p,q$ be distinct, full-support probability
vectors of equal length satisfying $p\prec q$.
Then
\begin{align}
    C^*(p\to q)
    =
    \left\lfloor 2^{\Delta_R(p,q)}\right\rfloor+1.
    \label{eq:main-result}
\end{align}
Equivalently, for every integer $m\geq 2$,
\begin{align}
    C^*(p\to q)\leq m
    \quad\Longleftrightarrow\quad
    \Delta_R(p,q)<\log_2m.
    \label{eq:main-equivalence}
\end{align}
The optimal cost is attained by a finite-dimensional
pure catalyst returned exactly and uncorrelated
with the target state.
\end{theorem}

The optimization places no upper bound or penalty on the
catalyst's Schmidt rank. Theorem~\ref{thm:catalytic-cost}
therefore characterizes the minimum message alphabet size with unrestricted
finite catalytic assistance. The proof is given in
Sec.~\ref{sec:proof-characterization}.

The strict inequality in
Eq.~\eqref{eq:main-equivalence} is essential:
when $2^{\Delta_R(p,q)}$ is an integer $M$,
the minimum communication alphabet has
$M+1$ elements. In particular, $C^*(p\to q)=2$ precisely
when $\Delta_R(p,q)<1$. In this case,
a binary communication alphabet is optimal,
so Alice needs to send Bob only one classical bit.

Before proving Theorem~\ref{thm:catalytic-cost}, we apply it to an
explicit family of transformations to show that catalytic assistance
can reduce the required message alphabet size by an arbitrarily
large factor.

\section{Maximal one-shot compression}
\label{sec:maximal-compression}

Catalytic assistance can reduce the minimum
message alphabet size from its dimension-dependent
maximum to two elements, corresponding to a single
classical bit. We establish this separation with an explicit
family of full-support Schmidt vectors.

The dimension bound~\eqref{eq:majorization-message-bound}
gives $\Bto(p\to q)\leq d$ for Schmidt vectors of length $d$.
For every integer $d\geq3$, define
\begin{align}
    p^{(d)}
    &=\left(
    \frac12,\frac14,
    \underbrace{\frac{1}{4(d-2)},\ldots,
    \frac{1}{4(d-2)}}_{d-2}
    \right),\nonumber\\
    q^{(d)}
    &=\left(
    \frac34,
    \underbrace{\frac{1}{4(d-1)},\ldots,
    \frac{1}{4(d-1)}}_{d-1}
    \right).
    \label{eq:maximal-compression-family}
\end{align}
Both vectors are normalized, strictly positive,
and arranged in nonincreasing order. We will show that
\begin{align}
    \boxed{
    \begin{aligned}
    B_{\rightarrow}(p^{(d)}\to q^{(d)})&=d,\\
    C^*(p^{(d)}\to q^{(d)})&=2.
    \end{aligned}
    }
    \label{eq:maximal-compression}
\end{align}

\emph{Noncatalytic cost.}
Write $P_d=\operatorname{diag}(p^{(d)})$ and
$Q_d=\operatorname{diag}(q^{(d)})$, and set
\begin{align}
    \beta_d:=\frac{1}{4(d-1)}.
\end{align}
The target reduced state has the form
\begin{align}
    Q_d=\beta_d I_d+
    \left(\frac34-\beta_d\right)
    |1\rangle\langle1|.
\end{align}
Every component of $p^{(d)}$ is strictly greater
than $\beta_d$. Therefore the numbers
\begin{align}
    w_i:=\frac{p_i^{(d)}-\beta_d}{3/4-\beta_d},
    \qquad 1\leq i\leq d,
\end{align}
are positive and sum to one. Choosing permutation
unitaries $V_i$ with $V_i|1\rangle=|i\rangle$ gives
\begin{align}
    P_d=\sum_{i=1}^{d}w_iV_iQ_dV_i^\dagger.
\end{align}
This proves both $p^{(d)}\prec q^{(d)}$ and the
existence of an exact protocol with a message alphabet of size $d$.

Conversely, any decomposition of $P_d$ into $M$
unitary conjugates of $Q_d$ implies
\begin{align}
    P_d-\beta_d I_d
    =\left(\frac34-\beta_d\right)
    \sum_{j=1}^{M}\lambda_j
    |v_j\rangle\langle v_j|,
    \label{eq:rank-obstruction}
\end{align}
where $\lambda_j>0$, $\sum_j\lambda_j=1$, and
$|v_j\rangle=U_j|1\rangle$.
The left-hand side is positive definite and has
rank $d$, whereas the right-hand side has rank at
most $M$. Thus $M\geq d$, and
\begin{align}
    B_{\rightarrow}(p^{(d)}\to q^{(d)})=d.
    \label{eq:maximal-noncatalytic-cost}
\end{align}

\emph{Catalytic cost.}
We prove that $\Delta_R(p^{(d)},q^{(d)})<1$ and
apply Theorem~\ref{thm:catalytic-cost}.

Let $u_n=(1/n,\ldots,1/n)$ denote the uniform
probability vector of length $n$, and let $0_d$
denote the zero vector of length $d$.
In particular, $u_2=(1/2,1/2)$.
The key observation is the auxiliary majorization
\begin{align}
    q^{(d)}\otimes u_2
    \prec p^{(d)}\oplus0_d.
    \label{eq:compression-auxiliary-majorization}
\end{align}
To verify it, decompose the two vectors, up to
permutations, as
\begin{align}
    q^{(d)}\otimes u_2
    &=\left(\frac38,\frac38\right)
    \oplus\frac14u_{2d-2},\nonumber\\
    p^{(d)}\oplus0_d
    &=\left(\frac12,\frac14\right)
    \oplus\frac14\left(u_{d-2}\oplus0_d\right).
\end{align}
The first block satisfies
$(3/8,3/8)\prec(1/2,1/4)$.
The second satisfies
$u_{2d-2}\prec u_{d-2}\oplus0_d$, since a uniform
probability vector is majorized by every probability
vector of the same length. Taking direct sums proves
Eq.~\eqref{eq:compression-auxiliary-majorization}.

The vectors in
Eq.~\eqref{eq:compression-auxiliary-majorization}
are not permutations of one another:
the left-hand side has $2d$ positive entries,
whereas the right-hand side has only $d$.
For every $0<\alpha<\infty$, strict Schur concavity
applied to
Eq.~\eqref{eq:compression-auxiliary-majorization}
gives
\begin{align}
    H_\alpha(q^{(d)}\otimes u_2)
    >H_\alpha(p^{(d)}\oplus0_d).
    \label{eq:compression-strict-renyi}
\end{align}
To express this inequality in terms of the original
source and target vectors, we use additivity:
\begin{align}
    H_\alpha(q^{(d)}\otimes u_2)
    &=H_\alpha(q^{(d)})+H_\alpha(u_2)\nonumber\\
    &=H_\alpha(q^{(d)})+1,
\end{align}
since $u_2=(1/2,1/2)$ has R\'enyi entropy one.
Moreover, appending zero entries does not change
the entropy, so
\begin{align}
    H_\alpha(p^{(d)}\oplus0_d)=H_\alpha(p^{(d)}).
\end{align}
Substituting these identities into
Eq.~\eqref{eq:compression-strict-renyi} yields
\begin{align}
    H_\alpha(p^{(d)})-H_\alpha(q^{(d)})<1.
\end{align}
At the endpoints,
\begin{align}
    H_0(p^{(d)})-H_0(q^{(d)})&=0,\nonumber\\
    H_\infty(p^{(d)})-H_\infty(q^{(d)})
    &=\log_2\frac32<1.
\end{align}
Continuity on the compactified interval
$[0,\infty]$ now yields
\begin{align}
    0<\Delta_R(p^{(d)},q^{(d)})<1,
\end{align}
where positivity follows from
$p^{(d)}\prec q^{(d)}$ and $p^{(d)}\neq q^{(d)}$.
Theorem~\ref{thm:catalytic-cost} consequently implies
\begin{align}
    C^*(p^{(d)}\to q^{(d)})=2.
    \label{eq:maximal-catalytic-cost}
\end{align}
The catalyst is finite dimensional and returned
exactly and uncorrelated with the target.

Combining the two costs establishes
Eq.~\eqref{eq:maximal-compression}. Their ratio is
$d/2$, which diverges as $d\to\infty$. In fixed-length binary encoding, the cost decreases
from $\lceil\log_2d\rceil$ bits to one bit.

\section{Proof of the catalytic characterization}
\label{sec:proof-characterization}

We prove Theorem~\ref{thm:catalytic-cost} by establishing
the two implications in Eq.~\eqref{eq:main-equivalence}
separately. We first show that any exact catalytic protocol with
a message alphabet of size at most $m$ obeys
$\Delta_R(p,q)<\log_2m$.
We then construct a finite pure catalyst permitting such a protocol
for every integer $m\geq2$ satisfying this inequality.

\subsection{A catalyst-independent lower bound}
\label{subsec:lower-bound}

Our first result shows that the Rényi entropy
difference provides a strict lower bound on
the logarithm of the message alphabet size.

\begin{proposition}[Catalyst-independent lower bound]
\label{prop:lower-bound}
Let $p,q$ be distinct, full-support probability
vectors of equal length satisfying $p\prec q$.
If an exact catalytic transformation
$p\otimes r\to q\otimes r$ can be implemented
with a message alphabet of size at most $m$, then
\begin{align}
    \Delta_R(p,q)<\log_2m.
    \label{eq:strict-lower-bound}
\end{align}
\end{proposition}

\begin{proof}
Let $R=\operatorname{diag}(r)$ be the reduced
state of an arbitrary finite pure catalyst.
By the equal-weight unitary-orbit
characterization~\cite{DaftuarHayden2005},
a protocol with a message alphabet of size at most $m$ implies
\begin{align}
    P\otimes R=\frac1m\sum_{j=1}^{m}X_j,
    \qquad
    X_j=U_j(Q\otimes R)U_j^\dagger.
\end{align}

The case $m=1$ is impossible. Indeed, it would imply $P\otimes R\sim Q\otimes R$ and hence, by additivity of the von Neumann entropy, $H_1(p)=H_1(q)$. This contradicts strict Schur concavity, since $p\prec q$ and $p\neq q$. Therefore $m\geq2$.

Let $h=\operatorname{rank}R$ and $N=dh$.
Each $X_j$ is an $N\times N$ positive definite
matrix with spectrum $q\otimes r$. Consider the $N\times mN$ block-row matrix
\begin{align}
    V=\frac{1}{\sqrt m}
    \begin{bmatrix}
        X_1^{1/2} & \cdots & X_m^{1/2}
    \end{bmatrix},
\end{align}
and its Gram matrix
\begin{align}
    G=V^\dagger V.
\end{align}
By construction,
\begin{align}
    VV^\dagger
    =\frac1m\sum_{j=1}^{m}X_j
    =P\otimes R.
\end{align}
Since $V^\dagger V$ and $VV^\dagger$
have the same nonzero eigenvalues,
the spectrum of $G$ consists of the
$N$ eigenvalues of $P\otimes R$,
together with $(m-1)N$ zeros. Thus,
up to ordering,
\begin{align}
    \lambda(G)
    =(p\otimes r)\oplus 0_{(m-1)N},
    \label{eq:gram-spectrum}
\end{align}
where $\lambda(X)$ denotes the vector
of eigenvalues of $X$, counted with
multiplicity.

We now regard $G$ as an $m\times m$
block matrix, with blocks of size
$N\times N$. Its $(i,j)$ block is
\begin{align}
    G_{ij}
    =\frac1m X_i^{1/2}X_j^{1/2}.
\end{align}
In particular, its diagonal blocks
are $G_{jj}=X_j/m$.

Let $\mathcal D$ denote the pinching map
that removes all off-diagonal blocks.
Explicitly,
\begin{align}
    \mathcal D(G)
    =\frac1m\bigoplus_{j=1}^{m}X_j.
    \label{eq:gram-pinching}
\end{align}
It is well known that pinching can be
expressed as an average of unitary
conjugations~\cite{Bhatia2000},
\begin{align}
    \mathcal D(G)
    =\frac1m\sum_{k=0}^{m-1}
    W^kG(W^\dagger)^k,
\end{align}
where
\begin{align}
    W=\operatorname{diag}
    (1,\omega,\ldots,\omega^{m-1})\otimes I_N,
    \qquad
    \omega=e^{2\pi i/m}.
\end{align}
It follows that
\begin{align}
    \lambda(\mathcal D(G))\prec\lambda(G).
\end{align}
Since each $X_j$ is unitarily equivalent
to $Q\otimes R$, we have
\begin{align}
    \lambda(\mathcal D(G))
    =u_m\otimes q\otimes r,
\end{align}
up to ordering. Combining this with
Eq.~\eqref{eq:gram-spectrum} yields
\begin{align}
    u_m\otimes q\otimes r
    \prec
    (p\otimes r)\oplus0_{(m-1)N}.
    \label{eq:pinching-majorization}
\end{align}

Rényi entropies are strictly Schur-concave
for every $0<\alpha<\infty$
\cite{MarshallOlkinArnold2011}. Therefore,
Eq.~\eqref{eq:pinching-majorization} implies
\begin{align}
    H_\alpha(u_m\otimes q\otimes r)
    >
    H_\alpha((p\otimes r)\oplus0_{(m-1)N}),
\end{align}
where the inequality is strict because
the two vectors cannot be permutations
of one another: the former has $mN$
strictly positive entries, while the
latter has only $N$, since $p$, $q$,
and $r$ have full support and $m\geq2$.

Using the additivity of Rényi entropies
under tensor products, and observing
that appending zeros does not change
$H_\alpha$ for $\alpha>0$, we obtain
\begin{align}
    \log_2m+H_\alpha(q)+H_\alpha(r)
    >
    H_\alpha(p)+H_\alpha(r).
\end{align}
The catalyst contribution cancels,
yielding
\begin{align}
    H_\alpha(p)-H_\alpha(q)
    <\log_2m,
    \qquad 0<\alpha<\infty.
    \label{eq:renyi-lower-bound-finite}
\end{align}

The endpoint $\alpha=\infty$ requires
a separate argument. Since all the
matrices $X_j$ are positive definite,
\begin{align}
    P\otimes R-\frac1mX_1
    =\frac1m\sum_{j=2}^{m}X_j>0.
\end{align}
Comparing the largest eigenvalues gives
\begin{align}
    p_1r_1>\frac{q_1r_1}{m},
\end{align}
and consequently
\begin{align}
    H_\infty(p)-H_\infty(q)<\log_2m.
\end{align}
At $\alpha=0$, the entropy difference
vanishes because $p$ and $q$ have
equal full support.

Finally, the Rényi entropy difference
is continuous on the compactified
interval $[0,\infty]$. Since it is
strictly smaller than $\log_2m$ at
every point, its maximum satisfies
\begin{align}
    \Delta_R(p,q)<\log_2m.
\end{align}
\end{proof}

Proposition~\ref{prop:lower-bound} implies
that every achievable catalytic alphabet
size must satisfy
\begin{align}
    m>2^{\Delta_R(p,q)}.
\end{align}
Since $m$ is an integer,
\begin{align}
    C^*(p\to q)
    \geq
    \left\lfloor
    2^{\Delta_R(p,q)}
    \right\rfloor+1.
    \label{eq:catalytic-lower-bound}
\end{align}

\subsection{Construction of a finite catalyst}
\label{subsec:finite-catalyst}

We now prove the converse implication in
Eq.~\eqref{eq:main-equivalence}. Fix an integer $m\geq2$
satisfying $\Delta_R(p,q)<\log_2m$. By the unitary-orbit
characterization, it suffices to construct a finite-dimensional
positive definite density matrix $R$ such that
\begin{align}
    P\otimes R\in\mathcal F_m(Q\otimes R).
    \label{eq:finite-catalyst-goal}
\end{align}
The construction proceeds in three steps:
\begin{enumerate}
    \item Proposition~\ref{prop:root-extraction} shows how an
    $m^k$-term catalytic decomposition on $k$ copies yields an
    $m$-term catalytic decomposition on one copy. This reduces
    our task to a finite multicopy construction.

    \item Proposition~\ref{prop:uniform-source} establishes the
    required catalytic decomposition when the source spectrum
    is uniform.

    \item For a general source, we decompose its tensor powers
    into uniform spectral blocks. Lemma~\ref{lem:spectral-allocation}
    distributes the target spectrum among these blocks. Finite
    rational realizations and a common catalyst then combine
    the block decompositions using a message alphabet of size at most $m^k$,
    allowing the first step to be applied.
\end{enumerate}

\subsubsection{Exact root extraction}
\label{subsec:root-extraction}

This procedure builds on the telescoping direct-sum construction of Duan et al.~\cite{DuanFengLiYing2005}, who established the connection between multicopy entanglement transformations and single-copy catalytic transformations. Our construction refines their approach by incorporating intermediate matrices obtained through successive averages over the digits of the outcome labels. This additional ingredient allows us to control the number of unitary conjugates, reducing an $m^k$-term decomposition on $k$ copies to an $m$-term catalytic decomposition on a single copy, as established in the following proposition.

\begin{proposition}[Exact root extraction]
\label{prop:root-extraction}
Let $P,Q$ be positive definite density
matrices of equal dimension, and let
$m\geq2$ be an integer.
Suppose that, for some integer $k\geq1$
and some finite-dimensional positive
definite density matrix $S$,
\begin{align}
    P^{\otimes k}\otimes S
    \in
    \mathcal F_{m^k}
    (Q^{\otimes k}\otimes S).
    \label{eq:root-assumption}
\end{align}
Then there exists a finite-dimensional
positive definite density matrix $R$
such that
\begin{align}
    P\otimes R\in\mathcal F_m(Q\otimes R).
    \label{eq:root-conclusion}
\end{align}
\end{proposition}

\begin{proof}
By assumption, there exist $m^k$ matrices
$Y_v$, indexed by $v\in[m]^k$, where
$[m]=\{0,\ldots,m-1\}$, such that
\begin{align}
    P^{\otimes k}\otimes S
    =
    \frac1{m^k}\sum_{v\in[m]^k}Y_v,
    \qquad
    Y_v\sim Q^{\otimes k}\otimes S.
    \label{eq:root-decomposition}
\end{align}
Here $\sim$ denotes unitary equivalence. 

The idea is to replace the average over
$m^k$ matrices by $k$ successive averages,
each involving only $m$ terms. To this
end, we regard the index $v$ as a word
of $k$ digits and average over one digit
at a time.

Consider the block-diagonal density matrix
\begin{align}
    Z=\frac1{m^k}
    \bigoplus_{v\in[m]^k}Y_v.
    \label{eq:root-Z}
\end{align}
Each block is labeled by a word
$v=(v_1,\ldots,v_k)\in[m]^k$.

For each $\ell=1,\ldots,k$, let $W_\ell$
be the unitary that cyclically shifts
the $\ell$th digit of the block labels,
leaving the remaining digits and the
internal matrix indices unchanged.
Define the averaging map
\begin{align}
    \mathcal T_\ell(X)
    :=
    \frac1m\sum_{j=0}^{m-1}
    W_\ell^j X(W_\ell^\dagger)^j.
    \label{eq:root-averaging-map}
\end{align}
When applied to a block-diagonal matrix,
$\mathcal T_\ell$ replaces each block
by the average of the $m$ blocks obtained
by varying its $\ell$th digit while
keeping the other digits fixed.

We apply these maps successively,
starting from the last digit, and define
\begin{align}
    \Sigma_\ell
    :=
    \mathcal T_{\ell+1}\circ\cdots
    \circ\mathcal T_k(Z),
    \qquad 0\leq\ell\leq k,
    \label{eq:root-sigma}
\end{align}
with $\Sigma_k=Z$.

After averaging the last $k-\ell$
digits, the blocks depend only on
the first $\ell$ digits. More explicitly,
the block of $\Sigma_\ell$ labeled
by $v=(v_1,\ldots,v_k)$ is
\begin{align}
    [\Sigma_\ell]_{v,v}
    =
    \frac1{m^k}
    \frac1{m^{k-\ell}}
    \sum_{w\in[m]^{k-\ell}}
    Y_{(v_1,\ldots,v_\ell,w)}.
    \label{eq:root-block-average}
\end{align}
Thus, blocks sharing the same first
$\ell$ digits are identical.

At $\ell=0$, all digits have been
averaged, so every block equals the
mean of the original matrices.
By Eq.~\eqref{eq:root-decomposition},
\begin{align}
    \Sigma_0
    =
    \frac{I_{m^k}}{m^k}
    \otimes P^{\otimes k}\otimes S.
    \label{eq:root-sigma-zero}
\end{align}
At the opposite endpoint, no averaging
has been performed. Since every $Y_v$
is unitarily equivalent to
$Q^{\otimes k}\otimes S$, we have
\begin{align}
    \Sigma_k=Z
    \sim
    \frac{I_{m^k}}{m^k}
    \otimes Q^{\otimes k}\otimes S.
    \label{eq:root-sigma-k}
\end{align}

Moreover, the recursive definition gives
\begin{align}
    \Sigma_\ell
    =
    \mathcal T_{\ell+1}(\Sigma_{\ell+1})
    =
    \frac1m\sum_{j=0}^{m-1}T_{\ell,j},
    \label{eq:root-chain}
\end{align}
where
\begin{align}
    T_{\ell,j}
    :=
    W_{\ell+1}^j\Sigma_{\ell+1}
    (W_{\ell+1}^\dagger)^j
    \sim\Sigma_{\ell+1}.
\end{align}
We have therefore obtained a chain of
intermediate matrices, in which each
step involves only $m$ unitary conjugates.

We now combine these intermediate
matrices into a single catalyst.
For $\ell=0,\ldots,k-1$, define
\begin{align}
    R_\ell
    :=
    P^{\otimes\ell}
    \otimes Q^{\otimes(k-1-\ell)}
    \otimes\Sigma_\ell,
\end{align}
and set
\begin{align}
    R=\frac1k\bigoplus_{\ell=0}^{k-1}R_\ell.
    \label{eq:root-catalyst}
\end{align}
Since each $R_\ell$ is a positive
definite density matrix of the same
dimension, $R$ is a finite-dimensional
positive definite density matrix.

For each $j=0,\ldots,m-1$, let
\begin{align}
    A_j:=\frac1k\bigoplus_{\ell=0}^{k-1}
    \bigl(P^{\otimes(\ell+1)}
    \otimes Q^{\otimes(k-1-\ell)}
    \otimes T_{\ell,j}\bigr).
    \label{eq:root-Aj}
\end{align}
Using Eq.~\eqref{eq:root-chain}, we obtain
\begin{align}
    \frac1m\sum_{j=0}^{m-1}A_j=P\otimes R.
    \label{eq:root-average}
\end{align}

It remains to show that
$A_j\sim Q\otimes R$.
For $0\leq\ell<k-1$, using
$T_{\ell,j}\sim\Sigma_{\ell+1}$,
we obtain
\begin{align}
    &P^{\otimes(\ell+1)}
    \otimes Q^{\otimes(k-1-\ell)}
    \otimes T_{\ell,j}
    \nonumber\\
    &\quad\sim
    Q\otimes
    \left(
    P^{\otimes(\ell+1)}
    \otimes Q^{\otimes(k-2-\ell)}
    \otimes\Sigma_{\ell+1}
    \right)
    \nonumber\\
    &\quad=Q\otimes R_{\ell+1},
    \label{eq:root-block-shift}
\end{align}
where tensor factors have been
rearranged by unitary permutations.

For the last block, the endpoint
relations
\eqref{eq:root-sigma-zero} and
\eqref{eq:root-sigma-k}, together with
$T_{k-1,j}\sim\Sigma_k$, give
\begin{align}
    P^{\otimes k}\otimes T_{k-1,j}
    &\sim P^{\otimes k}\otimes\Sigma_k
    \nonumber\\
    &\sim Q^{\otimes k}\otimes\Sigma_0
    \nonumber\\
    &=Q\otimes R_0,
    \label{eq:root-cycle-closure}
\end{align}
up to a permutation of tensor factors.

Thus the blocks of $A_j$ are unitarily
equivalent, in order, to
$Q\otimes R_1,\ldots,Q\otimes R_{k-1},
Q\otimes R_0$. Consequently,
\begin{align}
    A_j
    &\sim\frac1k
    \left(
    \bigoplus_{\ell=1}^{k-1}
    Q\otimes R_\ell
    \;\oplus\;
    Q\otimes R_0
    \right)
    \nonumber\\
    &\sim Q\otimes R,
\end{align}
where the last equivalence follows
by cyclically permuting the blocks.

Combining this with
Eq.~\eqref{eq:root-average}, we obtain
\begin{align}
    P\otimes R
    =\frac1m\sum_{j=0}^{m-1}A_j,
    \qquad A_j\sim Q\otimes R,
\end{align}
which proves
$P\otimes R\in\mathcal F_m(Q\otimes R)$.
\end{proof}

\subsubsection{Uniform sources}
\label{subsec:uniform-sources}

We now construct catalysts for
uniform source spectra. This special
case will serve as a building block
for the general construction.

Let $p=u_D$ and let $q$ be a
full-support probability vector of
dimension $D$, with largest component
$q_1$. Since all Rényi entropies of
$u_D$ equal $\log_2D$, and
$H_\alpha(q)$ is nonincreasing in
$\alpha$, we have
\begin{align}
    \Delta_R(u_D,q)
    =\log_2D-H_\infty(q)
    =\log_2(Dq_1).
    \label{eq:uniform-renyi-gap}
\end{align}
Thus the condition
$\Delta_R(u_D,q)<\log_2m$
reduces to
\begin{align}
    Dq_1<m.
    \label{eq:uniform-spectral-condition}
\end{align}
Under this spectral condition, the following proposition
provides an exact catalytic decomposition with $m$ terms.

\begin{proposition}[Uniform sources]
\label{prop:uniform-source}
Let $\sigma$ be a positive definite
density matrix of dimension $D$, and
let $m\geq2$ be an integer satisfying
\begin{align}
    D\lambda_{\max}(\sigma)<m.
\end{align}
Then there exists a finite-dimensional
positive definite density matrix
$R_\sigma$ such that
\begin{align}
    \frac{I_D}{D}\otimes R_\sigma
    \in
    \mathcal F_m(\sigma\otimes R_\sigma).
    \label{eq:uniform-catalytic-decomposition}
\end{align}
\end{proposition}

\begin{proof}
We first establish an exact multicopy
decomposition and then apply
Proposition~\ref{prop:root-extraction}
to obtain the desired one-copy
catalytic decomposition.

The case $\sigma=I_D/D$ is immediate.
Otherwise, let $q$ denote the spectrum
of $\sigma$, with
$q_1=\lambda_{\max}(\sigma)$.
We show that, for every sufficiently
large integer $b$,
\begin{align}
    \frac{I_{D^b}}{D^b}
    \in
    \mathcal F_{m^b}(\sigma^{\otimes b}).
    \label{eq:uniform-multicopy}
\end{align}

To this end, set
\begin{align}
    N=D^b,\qquad M=m^b,
    \qquad y=(q^{\otimes b})^\downarrow,
\end{align}
where the arrow denotes nonincreasing
rearrangement.

Proving Eq.~\eqref{eq:uniform-multicopy}
is equivalent to showing that there exist
Hermitian matrices $A_1,\ldots,A_M\in\mathbb C^{N\times N}$
such that
\begin{align}
    \spec(A_j)&=\frac{y}{M},
    \qquad 1\leq j\leq M,\nonumber\\
    \sum_{j=1}^{M}A_j&=\frac{I_N}{N}.
\end{align}

The additive Horn theorem~\cite[Theorem~17]{Fulton2000} implies that
proving the existence of these matrices is equivalent
to verifying the inequalities
\begin{align}
    \frac1M\sum_{j=1}^{M}
    \sum_{i\in I_j}y_i
    \geq\frac{r}{N},
    \label{eq:uniform-horn}
\end{align}
for every $1\leq r<N$ and every
admissible Horn tuple
$(I_1,\ldots,I_M,K)$ of $r$-element
index sets, namely those with
nonzero Littlewood--Richardson
coefficients.

We prove in Appendix~\ref{app:uniform} that these
inequalities are satisfied for all
sufficiently large $b$.
The strict hypothesis $Dq_1<m$
plays a crucial role: it implies
\begin{align}
    \frac{Ny_1}{M}
    =
    \left(\frac{Dq_1}{m}\right)^b
    \longrightarrow0.
\end{align}
Combined with concentration estimates
for the spectrum of $q^{\otimes b}$
and the combinatorial constraints on
admissible Horn tuples, this allows
us to verify the entire family of
inequalities in
Eq.~\eqref{eq:uniform-horn}.
The detailed argument is given
in Appendix~\ref{app:uniform}.

The additive Horn theorem therefore
establishes the exact decomposition
\eqref{eq:uniform-multicopy}
at a finite blocklength $b$.

Finally, applying
Proposition~\ref{prop:root-extraction}
with $P=I_D/D$, $Q=\sigma$,
$k=b$, and a trivial auxiliary state
$S$, we obtain a finite-dimensional
positive definite density matrix
$R_\sigma$ satisfying
\begin{align}
    \frac{I_D}{D}\otimes R_\sigma
    \in
    \mathcal F_m(\sigma\otimes R_\sigma),
\end{align}
as claimed.
\end{proof}

\subsubsection{From uniform sources to a general source}
\label{subsec:general-sources}

We now extend the construction to
arbitrary full-support probability vectors
$p\prec q$.

Fix an integer $m\geq2$ such that
\begin{align}
    \Delta_R(p,q)<\log_2m.
\end{align}
By Proposition~\ref{prop:root-extraction},
it suffices to construct, for some
finite integer $k$ and some
finite-dimensional auxiliary state $S$,
\begin{align}
    P^{\otimes k}\otimes S
    \in
    \mathcal F_{m^k}
    (Q^{\otimes k}\otimes S).
    \label{eq:general-multicopy-goal}
\end{align}

The case of a uniform source has already
been established in
Proposition~\ref{prop:uniform-source}.
We therefore assume that $p$ is nonuniform.

Our starting point is the decomposition
of tensor powers into uniform spectral
blocks, using the method of types
\cite{CoverThomas2006}.
For each type
$t=(t_1,\ldots,t_d)$, with
$\sum_i t_i=k$, define
\begin{align}
    z_t=\prod_{i=1}^d p_i^{t_i},
    \qquad
    N_t=\frac{k!}{\prod_{i=1}^d t_i!}.
\end{align}
Here $z_t$ is the eigenvalue associated
with type $t$, while $N_t$ is its
multiplicity. Thus, up to a permutation
of the basis,
\begin{align}
    P^{\otimes k}
    =\bigoplus_t z_t I_{N_t}.
    \label{eq:general-source-types}
\end{align}

Similarly, defining
\begin{align}
    w_s=\prod_{i=1}^d q_i^{s_i},
    \qquad
    N_s'=\frac{k!}{\prod_{i=1}^d s_i!},
\end{align}
we have
\begin{align}
    Q^{\otimes k}
    =\bigoplus_s w_s I_{N_s'}.
    \label{eq:general-target-types}
\end{align}
The number $T_k$ of possible source
types satisfies $T_k\leq(k+1)^d$.

Since every source block is proportional
to the identity, we would like to apply
Proposition~\ref{prop:uniform-source} to each
block separately.

To this end, we seek a partition of
the target spectrum into matrices
$Y_t$ such that
\begin{align}
    Q^{\otimes k}\sim\bigoplus_t Y_t,
    \label{eq:general-target-decomposition}
\end{align}
with each $Y_t$ corresponding to the
source block $z_t I_{N_t}$.

Let $n_{st}$ denote the number of
eigenvalues of target type $s$
assigned to the block $Y_t$.
The latter therefore takes the form
\begin{align}
    Y_t=\bigoplus_s w_s I_{n_{st}},
    \label{eq:general-target-blocks}
\end{align}
where the multiplicities $n_{st}$
must be nonnegative integers.

We now derive the conditions these
multiplicities must satisfy in order
to apply Proposition~\ref{prop:uniform-source}.

First, since $w_s$ occurs with
multiplicity $N_s'$ in the target
spectrum, every eigenvalue must be
assigned exactly once:
\begin{align}
    \sum_t n_{st}=N_s'.
    \label{eq:allocation-target-count}
\end{align}
Second, each target block must have
the same dimension as its corresponding
source block:
\begin{align}
    \sum_s n_{st}=N_t.
    \label{eq:allocation-source-count}
\end{align}
Finally, matching the traces of the
two blocks requires
\begin{align}
    \sum_s w_s n_{st}=N_tz_t.
    \label{eq:allocation-trace}
\end{align}

To apply Proposition~\ref{prop:uniform-source},
suppose that we allow $M$ terms in each
blockwise catalytic decomposition,
where the integer $M\geq2$ will be
specified below.

Under the preceding constraints,
the normalized matrix
\begin{align}
    \sigma_t=\frac{Y_t}{N_tz_t}
\end{align}
is a density matrix of dimension $N_t$.
The condition required by
Proposition~\ref{prop:uniform-source} is
\begin{align}
    N_t\lambda_{\max}(\sigma_t)<M,
\end{align}
which is equivalent to
\begin{align}
    n_{st}>0
    \quad\Longrightarrow\quad
    w_s<Mz_t.
    \label{eq:allocation-spectral-ceiling}
\end{align}

We have therefore reduced the problem
to finding nonnegative integer
multiplicities $n_{st}$ satisfying
the count, trace, and spectral
constraints above.

The following lemma provides an allocation satisfying
the required constraints when the variables $n_{st}$
are allowed to be nonnegative real numbers rather
than integers.

\begin{lemma}[Spectral allocation]
\label{lem:spectral-allocation}
For every $\rho>\Delta_R(p,q)$ and
all sufficiently large integers $k$,
there exist nonnegative real numbers
$n_{st}$ satisfying
\begin{align}
    \sum_t n_{st}&=N_s',
    &
    \sum_s n_{st}&=N_t,
    \label{eq:spectral-allocation-counts}\\
    \sum_s w_s n_{st}&=N_tz_t,
    \label{eq:spectral-allocation-trace}
\end{align}
and
\begin{align}
    n_{st}>0
    \quad\Longrightarrow\quad
    w_s\leq2^{k\rho}z_t.
    \label{eq:spectral-allocation-bound}
\end{align}
\end{lemma}

The proof of
Lemma~\ref{lem:spectral-allocation}
is given in Sec.~\ref{sec:allocation}.

We now show how to turn this real
allocation into a finite-dimensional
matrix construction. Choose parameters satisfying
\begin{align}
    \Delta_R(p,q)<\rho<\rho_1<\log_2m,
    \qquad
    M=\left\lceil2^{k\rho_1}\right\rceil.
    \label{eq:general-parameters}
\end{align}
For sufficiently large $k$, let
$n=(n_{st})$ be the allocation
provided by
Lemma~\ref{lem:spectral-allocation}.
Since $2^{k\rho}<M$, it satisfies
\begin{align}
    n_{st}>0
    \quad\Longrightarrow\quad
    w_s\leq2^{k\rho}z_t<Mz_t.
    \label{eq:general-strict-ceiling}
\end{align}

Consider the polytope
\begin{align}
\mathcal P=
\left\{
u\geq0:
\begin{array}{l}
\displaystyle\sum_t u_{st}=N_s',\\[2mm]
\displaystyle\sum_s u_{st}=N_t,\\[2mm]
u_{st}=0\text{ if }n_{st}=0
\end{array}
\right\}.
\label{eq:allocation-polytope}
\end{align}
This is a bounded rational polytope
containing the allocation $n$.
Only the count constraints and the allowed support
enter $\mathcal P$; the trace constraints are not
imposed on its individual points. This distinction
allows rational realizations even when the spectral
data are irrational.

Since $n$ has strictly positive
coordinates on every allowed pair,
it belongs to the relative interior
of $\mathcal P$.
Standard properties of rational polytopes
\cite{Ziegler1995} then allow us to write
\begin{align}
    n=\sum_{a=1}^{L}\theta_a n^{(a)},
    \qquad
    \theta_a\geq0,
    \qquad
    \sum_{a=1}^{L}\theta_a=1,
    \label{eq:allocation-rational-decomposition}
\end{align}
where the $n^{(a)}\in\mathcal P$
are rational points that can be
chosen arbitrarily close to $n$.

For each rational allocation
$n^{(a)}$, define its block means by
\begin{align}
    z_t^{(a)}
    :=
    \frac1{N_t}\sum_s w_s n_{st}^{(a)}.
    \label{eq:allocation-block-means}
\end{align}
The individual means $z_t^{(a)}$ need not equal $z_t$.
However, by linearity,
Eq.~\eqref{eq:allocation-rational-decomposition}
gives
\begin{align}
    z_t=\sum_{a=1}^{L}\theta_a z_t^{(a)}.
    \label{eq:allocation-convex-means}
\end{align}

Moreover, since the spectral
inequalities are strict at $n$,
choosing the rational points
sufficiently close to $n$ ensures
\begin{align}
    n_{st}^{(a)}>0
    \quad\Longrightarrow\quad
    w_s<Mz_t^{(a)}.
    \label{eq:allocation-rational-ceiling}
\end{align}

Finally, the vectors $z^{(a)}$
belong to an affine hyperplane
of dimension at most $T_k-1$, since
\begin{align}
    \sum_t N_tz_t^{(a)}=1.
\end{align}
By Carathéodory's theorem
\cite{Ziegler1995}, we may retain
at most $T_k$ of these rational
allocations, with new convex
weights, while preserving
Eq.~\eqref{eq:allocation-convex-means}.
Relabeling the retained allocations
and their weights, and using $L$ for
the size of the reduced family, we obtain
\begin{align}
    z_t=\sum_{a=1}^{L}\theta_a z_t^{(a)},
    \qquad
    \sum_{a=1}^{L}\theta_a=1,
    \qquad L\leq T_k.
    \label{eq:allocation-finite-means}
\end{align}
Each retained allocation satisfies
the count constraints and
Eq.~\eqref{eq:allocation-rational-ceiling}.

Let $h$ be a common denominator
of all the retained rational
multiplicities $n_{st}^{(a)}$,
so that $hn_{st}^{(a)}$ are
nonnegative integers. For each allocation $a$ and source
type $t$, define
\begin{align}
    Y_t^{(a)}
    =
    \bigoplus_s
    \frac{w_s}{h}I_{h n_{st}^{(a)}}.
    \label{eq:general-rational-blocks}
\end{align}
The count and trace constraints imply
\begin{align}
    \dim Y_t^{(a)}=hN_t,
    \qquad
    \operatorname{Tr}Y_t^{(a)}
    =N_tz_t^{(a)}.
    \label{eq:general-block-properties}
\end{align}

The introduction of $h$ amounts to
adjoining a uniform auxiliary state
$I_h/h$. Indeed, defining
\begin{align}
    X=P^{\otimes k}\otimes\frac{I_h}{h},
    \qquad
    Y=Q^{\otimes k}\otimes\frac{I_h}{h},
\end{align}
and
\begin{align}
    X^{(a)}
    =
    \bigoplus_t
    \frac{z_t^{(a)}}{h}I_{hN_t},
    \label{eq:general-Xa}
\end{align}
we have
\begin{align}
    X=\sum_{a=1}^{L}\theta_a X^{(a)},
    \qquad
    Y\sim\bigoplus_tY_t^{(a)}
    \quad\text{for every }a.
    \label{eq:general-convex-decomposition}
\end{align}

We can now apply
Proposition~\ref{prop:uniform-source}
to each pair of corresponding blocks.
For this purpose, normalize
\begin{align}
    \sigma_t^{(a)}
    :=
    \frac{Y_t^{(a)}}{N_tz_t^{(a)}}.
\end{align}
Since $\sigma_t^{(a)}$ has
dimension $hN_t$, the spectral
bound gives
\begin{align}
    hN_t\lambda_{\max}(\sigma_t^{(a)})
    =
    \max_{s:n_{st}^{(a)}>0}
    \frac{w_s}{z_t^{(a)}}
    <M.
    \label{eq:general-uniform-condition}
\end{align}

Hence Proposition~\ref{prop:uniform-source}
provides a finite-dimensional catalyst
$R_{a,t}$ and $M$ unitary conjugates
such that
\begin{align}
    \frac{z_t^{(a)}}{h}
    I_{hN_t}\otimes R_{a,t}
    =
    \frac1M\sum_{j=1}^{M}
    U_{a,t,j}
    \left(Y_t^{(a)}\otimes R_{a,t}\right)
    U_{a,t,j}^\dagger.
    \label{eq:general-block-catalysis}
\end{align}

To combine these blockwise
decompositions, we introduce a
common catalyst
\begin{align}
    R_{\mathrm{com}}
    :=
    \bigotimes_{a,t}R_{a,t}.
    \label{eq:general-common-catalyst}
\end{align}
For each pair $(a,t)$, let
$\widetilde U_{a,t,j}$ denote
the extension of $U_{a,t,j}$
to the common catalyst, acting
as the identity on all factors
other than $R_{a,t}$.
Explicitly,
\begin{align}
    \widetilde U_{a,t,j}
    =
    U_{a,t,j}\otimes I_{\mathrm{rest}},
\end{align}
up to a permutation of tensor factors,
where $I_{\mathrm{rest}}$ is the identity
on the remaining catalyst factors.

Tensoring
Eq.~\eqref{eq:general-block-catalysis}
with these remaining factors gives
\begin{align}
    &\frac{z_t^{(a)}}{h}
    I_{hN_t}\otimes R_{\mathrm{com}}
    \nonumber\\
    &\quad=
    \frac1M\sum_{j=1}^{M}
    \widetilde U_{a,t,j}
    (Y_t^{(a)}\otimes R_{\mathrm{com}})
    \widetilde U_{a,t,j}^{\dagger}.
    \label{eq:general-common-block}
\end{align}

For each fixed allocation $a$,
we can now take the direct sum
over all source types $t$, using the
direct-sum rule~\eqref{eq:direct-sum-budget}.
Using
Eq.~\eqref{eq:general-convex-decomposition},
we obtain
\begin{align}
    X^{(a)}\otimes R_{\mathrm{com}}
    =
    \frac1M\sum_{j=1}^{M}
    W_{a,j}
    (Y\otimes R_{\mathrm{com}})
    W_{a,j}^{\dagger},
    \label{eq:general-assignment-decomposition}
\end{align}
where the unitaries $W_{a,j}$ are
obtained by taking direct sums of
the extended block unitaries and
incorporating the permutation that
identifies $Y$ with
$\bigoplus_tY_t^{(a)}$.

Finally, recalling that
$X=\sum_{a=1}^{L}\theta_a X^{(a)}$,
we obtain
\begin{align}
    X\otimes R_{\mathrm{com}}
    =
    \sum_{a=1}^{L}\sum_{j=1}^{M}
    \frac{\theta_a}{M}
    W_{a,j}
    (Y\otimes R_{\mathrm{com}})
    W_{a,j}^{\dagger}.
    \label{eq:general-final-decomposition}
\end{align}
This is a convex combination of
at most $LM$ unitary conjugates.

We now verify that this number fits
within the desired budget $m^k$.
Since
\begin{align}
    L\leq T_k\leq(k+1)^d,
    \qquad
    M=\left\lceil2^{k\rho_1}\right\rceil,
\end{align}
we have
\begin{align}
    \frac{LM}{m^k}
    \leq
    2(k+1)^d
    2^{-k(\log_2m-\rho_1)}
    \longrightarrow0.
    \label{eq:general-message-budget}
\end{align}
Thus, for sufficiently large finite
$k$, we have $LM\leq m^k$.

By Corollary 2.7 of
Daftuar and Hayden~\cite{DaftuarHayden2005},
the convex decomposition
\eqref{eq:general-final-decomposition},
which contains at most $m^k$ terms,
can be rewritten as an equal-weight
average of exactly $m^k$
unitary conjugates.
Hence
\begin{align}
    X\otimes R_{\mathrm{com}}
    \in
    \mathcal F_{m^k}
    (Y\otimes R_{\mathrm{com}}).
\end{align}

Recalling the definitions of $X$
and $Y$, this is precisely
\begin{align}
    P^{\otimes k}\otimes S
    \in
    \mathcal F_{m^k}
    (Q^{\otimes k}\otimes S),
    \label{eq:general-multicopy-established}
\end{align}
with the finite-dimensional
auxiliary state
\begin{align}
    S=\frac{I_h}{h}\otimes R_{\mathrm{com}}.
\end{align}

We can finally apply
Proposition~\ref{prop:root-extraction},
which yields a finite-dimensional
positive definite density matrix $R$
such that
\begin{align}
    P\otimes R\in\mathcal F_m(Q\otimes R).
\end{align}

We have therefore shown that $\Cst(p\to q)\leq m$ for
every integer $m$ satisfying
$\Delta_R(p,q)<\log_2m$.
Together with the lower bound
derived in
Proposition~\ref{prop:lower-bound},
we obtain
\begin{align}
    C^*(p\to q)
    =
    \left\lfloor
    2^{\Delta_R(p,q)}
    \right\rfloor+1,
\end{align}
completing the proof of
Theorem~\ref{thm:catalytic-cost}.

\section{Proof of the spectral allocation lemma}
\label{sec:allocation}

We now prove Lemma~\ref{lem:spectral-allocation},
which provides the key ingredient for extending
the catalytic construction to general sources.

Recall that our goal is to distribute the
eigenvalues of $Q^{\otimes k}$ among the
uniform spectral blocks of $P^{\otimes k}$,
preserving the dimension and trace of each
block. In addition, a target eigenvalue
$w_s$ may be assigned to a source block
with eigenvalue $z_t$ only if
\begin{align}
    w_s\leq 2^{k\rho}z_t,
\end{align}
where $\rho>\Delta_R(p,q)$.

Although majorization guarantees the
existence of an allocation preserving
the dimension and trace constraints,
it does not, by itself, ensure this
additional spectral bound.

Our proof proceeds in two stages.
We first establish a necessary and
sufficient criterion for the existence
of an allocation satisfying a prescribed
spectral ceiling, expressed in terms
of inequalities for convex functions.
We then apply this criterion to tensor
powers and show that the required
inequalities hold for sufficiently large
$k$, using probabilistic estimates
for the tails of sums of independent
random variables.

\subsection{A finite allocation criterion}
\label{subsec:finite-allocation}

We first consider the allocation problem
in finite dimension, independently of
the tensor-power structure.

Let $x=(x_i)_{i=1}^N$ and
$y=(y_j)_{j=1}^N$ be strictly positive
probability vectors satisfying $x\prec y$.
An allocation is described by a
nonnegative matrix $\tau=(\tau_{ij})$,
where $\tau_{ij}$ represents the fraction
of the target entry $y_j$ assigned
to the source entry $x_i$.

We require
\begin{align}
\begin{aligned}
    \sum_j \tau_{ij}&=1,
    &
    \sum_i \tau_{ij}&=1,\\
    \sum_j \tau_{ij}y_j&=x_i.
\end{aligned}
\label{eq:allocation-constraints}
\end{align}
Thus $\tau$ is doubly stochastic, and
the weighted average assigned to
each source entry equals $x_i$.

For a fixed $\Gamma>1$, we seek an allocation
with an additional restriction: a target entry
$y_j$ may contribute to the source entry $x_i$
only if $y_j\leq\Gamma x_i$.
We therefore require
\begin{align}
    \tau_{ij}=0
    \quad\text{whenever}\quad
    y_j>\Gamma x_i.
    \label{eq:ceiling-assignment}
\end{align}
We call this additional restriction a
\emph{spectral ceiling}.

The ceiling constraint suggests comparing the source
entries with $x_i\leq b$ to the target entries with
$y_j\leq\Gamma b$, since the former can receive
contributions only from the latter.
We compare these two groups through their numbers
of entries and their total excess above a level $a$.
To this end, for $b>0$ and $a\in\mathbb R$, define
\begin{align}
    N_x(b)&:=\#\{i:x_i\leq b\},\nonumber\\
    J_x(a,b)&:=\sum_{x_i\leq b}(x_i-a)_+,
\end{align}
with analogous definitions for $y$.
The following lemma shows that comparisons of these
quantities give an exact criterion for the existence
of an allocation satisfying the prescribed ceiling.

\begin{lemma}[Scalar allocation criterion]
\label{lem:cumulative}
\label{lem:scalar-allocation}
Let $x,y$ be strictly positive probability vectors
of equal length, with $x\prec y$, and let $\Gamma>1$.
There exists a nonnegative matrix $\tau=(\tau_{ij})$
satisfying Eqs.~\eqref{eq:allocation-constraints}
and~\eqref{eq:ceiling-assignment} if and only if
\begin{align}
    N_x(b)&\leq N_y(\Gamma b),
    \label{eq:cum-count}\\
    J_x(a,b)&\leq J_y(a,\Gamma b),
    \qquad 0\leq a\leq b,
    \label{eq:cum-call}
\end{align}
for every value $b$ attained by $x$.
\end{lemma}

We call Eqs.~\eqref{eq:cum-count}
and~\eqref{eq:cum-call} the \emph{count inequalities}
and \emph{call inequalities}, respectively.
The functions $(u-a)_+$ and $(a-u)_+$ are called
call and put functions.

\begin{proof}
Suppose first that a matrix $\tau=(\tau_{ij})$
satisfies the requirements of the lemma.
Fix a value $b$ attained by $x$.
Since each row has unit sum and prescribed
mean $x_i$, Jensen's inequality gives
\begin{align*}
    f(x_i)\leq\sum_j\tau_{ij}f(y_j)
\end{align*}
for every nonnegative convex function
$f:\mathbb R\to\mathbb R$.
Summing over the source entries with $x_i\leq b$,
we obtain
\begin{align*}
    \sum_{x_i\leq b}f(x_i)
    &\leq\sum_{x_i\leq b}\sum_j\tau_{ij}f(y_j)\\
    &=\sum_{y_j\leq\Gamma b}
        f(y_j)\sum_{x_i\leq b}\tau_{ij}\\
    &\leq\sum_{y_j\leq\Gamma b}f(y_j).
\end{align*}
The equality uses the ceiling constraint:
if $x_i\leq b$ and $y_j>\Gamma b$, then
$y_j>\Gamma x_i$ and hence $\tau_{ij}=0$.
The last inequality follows from $f(y_j)\geq0$
and
\begin{align*}
    \sum_{x_i\leq b}\tau_{ij}
    \leq\sum_i \tau_{ij}=1.
\end{align*}
Choosing $f(u)=1$ and $f(u)=(u-a)_+$,
with $0\leq a\leq b$, yields
Eqs.~\eqref{eq:cum-count}
and~\eqref{eq:cum-call}, respectively.

For the converse, assume that
Eqs.~\eqref{eq:cum-count} and~\eqref{eq:cum-call}
hold. Arrange both vectors in nondecreasing order.
Starting from the zero matrix, we construct $\tau$
one row at a time, using an interval construction
similar to that of Beiglb\"ock and Juillet
\cite{BeiglbockJuillet2016}.

Suppose that the first $k$ rows have unit sum,
prescribed means, and satisfy the ceiling
constraint, and that
\begin{align}
    c_j^{(k)}:=1-\sum_{i=1}^{k}\tau_{ij}\geq0.
\end{align}
The quantity $c_j^{(k)}$ is the amount that
remains to be assigned to column $j$ in the
subsequent rows so that its final sum equals
one. We call these quantities the residual
capacities.

Assume also that
\begin{align}\label{eq:residual-call}
    \sum_{\substack{i>k\\x_i\leq b}}(x_i-a)_+
    \leq
    \sum_{y_j\leq\Gamma b}
    c_j^{(k)}(y_j-a)_+
\end{align}
for every $b\in\{x_{k+1},\dots,x_N\}$ and
$a\geq0$.
At $k=0$, these inequalities follow from
Eq.~\eqref{eq:cum-call}, since $c_j^{(0)}=1$;
for $a>b$, the left-hand side vanishes.
We construct row $k+1$ and then verify
Eq.~\eqref{eq:residual-call} with $k$ replaced
by $k+1$.

\medskip
\noindent\emph{Construction of the next row.}
Let $C_k:=N_y(\Gamma x_{k+1})$.
The allowed columns are $j=1,\dots,C_k$.
Associate to each of them the interval
\begin{align*}
    I_j^{(k)}:=
    \left[
        \sum_{t=1}^{j-1}c_t^{(k)},
        \sum_{t=1}^{j}c_t^{(k)}
    \right).
\end{align*}
Since $x_i\leq x_{k+1}$ for $i\leq k$,
every completed row uses only these columns.
Each such row has consumed one unit of capacity,
so
\begin{align*}
    \sum_{j=1}^{C_k}c_j^{(k)}=C_k-k.
\end{align*}
Thus the consecutive intervals $I_j^{(k)}$
cover $[0,C_k-k)$.
The count inequality gives
\begin{align*}
    C_k\geq N_x(x_{k+1})\geq k+1,
\end{align*}
so this interval has length at least one.

For $0\leq s\leq C_k-k-1$, define
\begin{align*}
    \mu_k(s):=
    \sum_{j=1}^{C_k}
    y_j|I_j^{(k)}\cap[s,s+1)|,
\end{align*}
where $|\cdot|$ denotes interval length.
We first show that
\begin{align*}
    \mu_k(0)\leq x_{k+1}\leq \mu_k(C_k-k-1).
\end{align*}

The completed rows imply
\begin{align*}
    \sum_jc_j^{(k)}&=N-k,\\
    \sum_jc_j^{(k)}y_j&=\sum_{i>k}x_i.
\end{align*}
Moreover, the count inequality at $b=x_N$
gives $N_y(\Gamma x_N)=N$, so
Eq.~\eqref{eq:residual-call} at $b=x_N$
involves all target entries.
Using the two identities above and
$(a-t)_+=(t-a)_++a-t$, we therefore obtain
\begin{align}
    \sum_{i>k}(a-x_i)_+
    \leq
    \sum_jc_j^{(k)}(a-y_j)_+.
    \label{eq:residual-put}
\end{align}

Among the allowed columns, let $v$ be the largest
target value $y_j$ such that
$|I_j^{(k)}\cap[0,1)|>0$, and let $w$ be the
smallest target value $y_j$ such that
$|I_j^{(k)}\cap[C_k-k-1,C_k-k)|>0$.
Both exist because the corresponding unit
intervals are covered by the $I_j^{(k)}$.
Applying Eq.~\eqref{eq:residual-put} at $a=v$,
and Eq.~\eqref{eq:residual-call} at
$b=x_{k+1}$ and $a=w$, gives 
\begin{align}
    v-x_{k+1}
    &\leq(v-x_{k+1})_+
    \nonumber\\
    &\leq\sum_{j=1}^{C_k}c_j^{(k)}(v-y_j)_+
    \nonumber\\
    &=\sum_{y_j<v}c_j^{(k)}(v-y_j)
    \nonumber\\
    &=v-\mu_k(0),
    \label{eq:window-mean-lower}\\[1ex]
    x_{k+1}-w
    &\leq(x_{k+1}-w)_+
    \nonumber\\
    &\leq\sum_{j=1}^{C_k}c_j^{(k)}(y_j-w)_+
    \nonumber\\
    &=\sum_{\substack{j\leq C_k\\y_j>w}}
        c_j^{(k)}(y_j-w)
    \nonumber\\
    &=\mu_k(C_k-k-1)-w.
    \label{eq:window-mean-upper}
\end{align}
In Eq.~\eqref{eq:window-mean-lower}, columns
$j>C_k$ contribute zero because
$y_j>\Gamma x_{k+1}\geq v$.
For $y_j<v$, the ordering gives
$I_j^{(k)}\subseteq[0,1)$, so
$c_j^{(k)}=|I_j^{(k)}\cap[0,1)|$.
Likewise, for $j\leq C_k$ with $y_j>w$,
\[
    c_j^{(k)}
    =|I_j^{(k)}\cap[C_k-k-1,C_k-k)|.
\]
The final equalities in
Eqs.~\eqref{eq:window-mean-lower}
and~\eqref{eq:window-mean-upper} follow from
the unit lengths and the definition of $\mu_k$,
since the contributions at $v$ and $w$ vanish.

Each intersection length defining $\mu_k(s)$
is continuous in $s$.
The endpoint bounds therefore give a position
$s_k\in[0,C_k-k-1]$ with
$\mu_k(s_k)=x_{k+1}$.
Define
\begin{align*}
    \tau_{k+1,j}:=
    \begin{cases}
        |I_j^{(k)}\cap[s_k,s_k+1)|,
        &j\leq C_k,\\
        0,&j>C_k.
    \end{cases}
\end{align*}
This row has unit sum, prescribed mean
$x_{k+1}$, and satisfies the ceiling constraint.
Also, $0\leq \tau_{k+1,j}\leq c_j^{(k)}$, so the
updated capacities
\begin{align*}
    c_j^{(k+1)}=c_j^{(k)}-\tau_{k+1,j}
\end{align*}
remain nonnegative.

\medskip
\noindent\emph{Preservation of the inequalities.}
Among the allowed columns, let $\ell$ and $r$
be, respectively, the smallest and largest
target values $y_j$ such that
\[
    |I_j^{(k)}\cap[s_k,s_k+1)|>0.
\]
Since the new row has unit sum and mean
$x_{k+1}$, we have $\ell\leq x_{k+1}\leq r$.
For $\ell<y_j<r$, the ordering of the intervals
implies $I_j^{(k)}\subseteq[s_k,s_k+1)$.
Hence
\begin{align*}
    c_j^{(k+1)}
    &=c_j^{(k)}-\tau_{k+1,j}\\
    &=|I_j^{(k)}|
      -|I_j^{(k)}\cap[s_k,s_k+1)|\\
    &=0.
\end{align*}

Fix $b\in\{x_i:i>k+1\}$.
Every column used by the new row satisfies
$y_j\leq\Gamma x_{k+1}\leq\Gamma b$.
We must show that
\begin{align}
    \sum_{\substack{i>k+1\\x_i\leq b}}(x_i-a)_+
    \leq
    \sum_{y_j\leq\Gamma b}
    c_j^{(k+1)}(y_j-a)_+
    \label{eq:residual-call-next}
\end{align}
for every $a\geq0$.

For $a\leq\ell$, every target value used by
the new row satisfies $y_j\geq a$.
The unit sum and prescribed mean give
\begin{align}
    \sum_j\tau_{k+1,j}(y_j-a)_+
    &=\sum_j\tau_{k+1,j}(y_j-a)
    \nonumber\\
    &=\sum_j\tau_{k+1,j}y_j
      -a\sum_j\tau_{k+1,j}
    \nonumber\\
    &=x_{k+1}-a
    \nonumber\\
    &=(x_{k+1}-a)_+.
    \label{eq:row-call-equality}
\end{align}
For $a\geq r$, all target values used by the
row and their mean $x_{k+1}$ are at most $a$,
so both sides of Eq.~\eqref{eq:row-call-equality}
vanish.
Thus this equality holds whenever
$a\leq\ell$ or $a\geq r$.

For these values of $a$, subtracting the
contribution of the new row from
Eq.~\eqref{eq:residual-call} gives
\begin{align*}
    \sum_{\substack{i>k+1\\x_i\leq b}}(x_i-a)_+
    &=\sum_{\substack{i>k\\x_i\leq b}}(x_i-a)_+\\
    &\quad-(x_{k+1}-a)_+\\
    &\leq\sum_{y_j\leq\Gamma b}
        c_j^{(k)}(y_j-a)_+\\
    &\quad-\sum_{y_j\leq\Gamma b}
        \tau_{k+1,j}(y_j-a)_+\\
    &=\sum_{y_j\leq\Gamma b}
        c_j^{(k+1)}(y_j-a)_+.
\end{align*}
This proves Eq.~\eqref{eq:residual-call-next}
outside $(\ell,r)$, including both endpoints.

On $[\ell,r]$, the right-hand side of
Eq.~\eqref{eq:residual-call-next} is affine
in $a$, because no residual capacity remains
at target values strictly between $\ell$ and $r$.
The left-hand side is convex in $a$, being a sum
of functions $a\mapsto(x_i-a)_+$.
It therefore lies below the line joining its
endpoint values, which is bounded above by
the affine right-hand side.
Hence Eq.~\eqref{eq:residual-call-next} holds
throughout $[\ell,r]$ as well.
If $\ell=r$, the preceding cases already cover
every $a$.
This completes the induction step.

\medskip
After all $N$ rows have been constructed,
each row has unit sum and prescribed mean,
and the ceiling constraint holds.
Every column sum is at most one.
Since the total sum of the matrix entries is $N$,
every column sum must equal one.
Undoing the initial permutations gives the
required allocation matrix.
\end{proof}

\subsection{Verification on tensor-power spectra}
\label{subsec:tensor-power-allocation}

We now prove Lemma~\ref{lem:spectral-allocation}
by verifying the scalar inequalities of
Lemma~\ref{lem:scalar-allocation}
for sufficiently large tensor powers.

Choose
\begin{align}
    \Delta_R(p,q)<\rho_0<\rho,
\end{align}
and, for each integer $k$, set
\begin{align}
    x=p^{\otimes k},
    \qquad
    y=q^{\otimes k},
    \qquad
    T=2^{k\rho_0},
    \qquad
    \Gamma=2^{k\rho}.
    \label{eq:allocation-parameters}
\end{align}
Since $p\prec q$, we also have
$x\prec y$.

If $p$ is uniform, all source types have the
same eigenvalue $z_t=d^{-k}$. The allocation
\begin{align}
    n_{st}=\frac{N_s'N_t}{d^k}
\end{align}
satisfies the count and trace constraints directly.
The spectral ceiling follows from
$\Delta_R(u_d,q)=\log_2(dq_1)<\rho$.
We therefore assume that $p$ is
nonuniform.

\paragraph{The count inequalities.}

Since $T<\Gamma$, it suffices to prove the stronger inequalities
$N_x(b)\leq N_y(Tb)$ for every $b>0$.
Let $I$ be uniformly distributed over $\{1,\ldots,d\}$, and define
\begin{align}
    A=\log_2q_I-\rho_0,
    \qquad
    B=\log_2p_I.
    \label{eq:count-random-variables}
\end{align}
Let $A_1,A_2,\ldots$ and $B_1,B_2,\ldots$ be independent copies
of these variables, respectively, and write
\begin{align}
\begin{aligned}
    \overline A_k&:=\frac1k\sum_{\ell=1}^k A_\ell,\\
    \overline B_k&:=\frac1k\sum_{\ell=1}^k B_\ell.
\end{aligned}
\end{align}
Each tuple of $k$ indices has probability $d^{-k}$.
Thus $2^{k\overline B_k}$ samples the entries of $x$ uniformly,
while $2^{k\overline A_k}$ samples the entries of $y/T$ uniformly.
For $v=k^{-1}\log_2b$, this gives
\begin{align}
\begin{aligned}
    \frac{N_x(b)}{d^k}&=\Pr(\overline B_k\leq v),\\
    \frac{N_y(Tb)}{d^k}&=\Pr(\overline A_k\leq v).
\end{aligned}
\label{eq:count-probability-representation}
\end{align}
Our goal is therefore to show that
\begin{align}
    \Pr(\overline B_k\leq v)\leq\Pr(\overline A_k\leq v)
\end{align}
for every $v\in\mathbb R$ and all sufficiently large $k$.
We use the following comparison lemma for averages of
independent random variables.

\begin{lemma}[Uniform tail comparison]
\label{lem:uniform-tail-comparison}
Let $A$ and $B$ be nonconstant
real-valued random variables with
finite support. Suppose that
\begin{align}
    \mathbb E[2^{\lambda A}]
    &<\mathbb E[2^{\lambda B}],
    &&\lambda>0,
    \label{eq:tail-moment-positive}\\
    \mathbb E[2^{\lambda A}]
    &>\mathbb E[2^{\lambda B}],
    &&\lambda<0,
    \label{eq:tail-moment-negative}
\end{align}
and that
\begin{align}
\begin{aligned}
    \mathbb E[A]&<\mathbb E[B],\\
    \min A&<\min B,\\
    \max A&<\max B.
\end{aligned}
    \label{eq:tail-endpoint-conditions}
\end{align}

Let $\overline A_k$ and $\overline B_k$ denote the averages
of $k$ independent copies of $A$ and $B$, respectively. For any
\begin{align}
    \mathbb E[A]<c<\mathbb E[B],
\end{align}
there exists an integer $k_0$ such
that, for every $k\geq k_0$,
\begin{align}
    \Pr(\overline B_k\leq v)
    &\leq
    \frac14\Pr(\overline A_k\leq v),
    &&v\leq c,
    \label{eq:uniform-lower-tail}\\
    \Pr(\overline A_k\geq v)
    &\leq
    \frac14\Pr(\overline B_k\geq v),
    &&v\geq c.
    \label{eq:uniform-upper-tail}
\end{align}
Both inequalities hold simultaneously
for all thresholds $v$ in the
indicated ranges.
\end{lemma}

The proof is given in Appendix~\ref{app:tails}
and adapts the large-deviation argument in
the proof of Proposition~2.5 of Aubrun and
Nechita~\cite{AubrunNechita2009} to obtain
the uniform multiplicative bounds stated above. We now verify its hypotheses for the variables in
Eq.~\eqref{eq:count-random-variables}.
Since $p$ is nonuniform and $p\prec q$, both $p$ and $q$
are nonuniform. Hence $A$ and $B$ are nonconstant and have finite support.

For every $\lambda\neq0$,
\begin{align}
    \frac{\mathbb E[2^{\lambda A}]}
         {\mathbb E[2^{\lambda B}]}
    =
    2^{-\lambda\rho_0}
    \frac{\sum_iq_i^\lambda}
         {\sum_ip_i^\lambda}.
    \label{eq:allocation-moment-ratio}
\end{align}

For $\lambda>1$, taking logarithms gives
\begin{align}
    &\log_2
    \frac{\mathbb E[2^{\lambda A}]}
         {\mathbb E[2^{\lambda B}]}=
    (\lambda-1)
    [H_\lambda(p)-H_\lambda(q)]
    -\lambda\rho_0<0,
\end{align}
where the strict inequality follows
from $\Delta_R(p,q)<\rho_0$.

For $0<\lambda<1$, concavity of
$u^\lambda$ and $p\prec q$ imply
\begin{align}
    \sum_iq_i^\lambda
    \leq\sum_ip_i^\lambda,
\end{align}
so the ratio in
Eq.~\eqref{eq:allocation-moment-ratio}
is strictly smaller than one.
The same conclusion holds at
$\lambda=1$.

For $\lambda<0$, convexity of
$u^\lambda$ instead gives
\begin{align}
    \sum_iq_i^\lambda
    \geq\sum_ip_i^\lambda,
\end{align}
and hence the moment ratio is
strictly greater than one.

Furthermore, concavity of the
logarithm and $p\prec q$ imply
\begin{align}
    \mathbb E[A]<\mathbb E[B].
\end{align}
The endpoint conditions follow from
\begin{align}
    q_d\leq p_d,
    \qquad
    \log_2\frac{q_1}{p_1}
    \leq\Delta_R(p,q)<\rho_0,
\end{align}
which yield
\begin{align}
    \min A<\min B,
    \qquad
    \max A<\max B.
\end{align}

Thus all hypotheses of Lemma~\ref{lem:uniform-tail-comparison} are satisfied. Choose the separator
\begin{align}
    c_{\mathrm{cnt}}:=\frac{\mathbb E[A]+\mathbb E[B]}{2},
\end{align}
which satisfies $\mathbb E[A]<c_{\mathrm{cnt}}<\mathbb E[B]$.

For $v\leq c_{\mathrm{cnt}}$, the lower-tail estimate gives
\begin{align}
    \Pr(\overline B_k\leq v)
    \leq\frac14\Pr(\overline A_k\leq v)
    \leq\Pr(\overline A_k\leq v).
\end{align}

For $v>c_{\mathrm{cnt}}$, the upper-tail estimate applied at $v+\varepsilon$, with $\varepsilon>0$, yields
\begin{align}
    \Pr(\overline A_k\geq v+\varepsilon)
    \leq\frac14\Pr(\overline B_k\geq v+\varepsilon).
\end{align}
Letting $\varepsilon\downarrow0$ and using continuity of probability from below, we obtain
\begin{align}
    \Pr(\overline A_k>v)
    \leq\frac14\Pr(\overline B_k>v).
\end{align}
Taking complements gives
\begin{align}
    \Pr(\overline B_k\leq v)
    \leq\Pr(\overline A_k\leq v).
\end{align}

Combining the two cases, we conclude that, for all sufficiently large $k$,
\begin{align}
    \Pr(\overline B_k\leq v)
    \leq\Pr(\overline A_k\leq v),
    \qquad v\in\mathbb R.
\end{align}

Using Eq.~\eqref{eq:count-probability-representation},
this stochastic comparison gives
\begin{align}
    N_x(b)\leq N_y(Tb)
    \leq N_y(\Gamma b),
\end{align}
where the last inequality follows
from $T<\Gamma$ and the monotonicity
of the counting function.

This proves the count inequalities~\eqref{eq:cum-count}.

\paragraph{The call inequalities.}

We now verify Eq.~\eqref{eq:cum-call}.
To control the call sums, we compare the total spectral mass
carried by suitable intervals. This leads to sampling indices
according to the entries of $p$ and $q$.
Let $I_p$ and $I_q$ satisfy
\begin{align}
    \Pr(I_p=i)=p_i,
    \qquad
    \Pr(I_q=j)=q_j,
\end{align}
and define
\begin{align}
    Z_p:=\log_2p_{I_p},
    \qquad
    Z_q:=\log_2q_{I_q}.
    \label{eq:mass-random-variables}
\end{align}
For any probability vector $z$, define the lower and upper spectral masses
\begin{align}
\begin{aligned}
    F_z(s)&:=\sum_{z_i\leq s}z_i,\\
    G_z(s)&:=\sum_{z_i\geq s}z_i.
\end{aligned}
\label{eq:spectral-mass-functions}
\end{align}
A tuple $(i_1,\ldots,i_k)$ of independent indices sampled according to $p$
has probability $p_{i_1}\cdots p_{i_k}$, equal to the corresponding
entry of $x$; the same holds for $q$ and $y$.
Consequently, writing $\overline Z_{p,k}$ and $\overline Z_{q,k}$
for the averages of $k$ independent copies, we have,
for $v=k^{-1}\log_2s$,
\begin{align}
    F_x(s)
    &=\Pr(\overline Z_{p,k}\leq v),
    \nonumber\\
    F_y(Ts)
    &=\Pr(\overline Z_{q,k}-\rho_0\leq v),
    \nonumber\\
    G_x(s)
    &=\Pr(\overline Z_{p,k}\geq v),
    \nonumber\\
    G_y(Ts)
    &=\Pr(\overline Z_{q,k}-\rho_0\geq v).
    \label{eq:spectral-mass-probabilities}
\end{align}

Choose a constant $c$ satisfying
\begin{align}
    -H_1(q)-\rho_0<c<-H_1(p).
    \label{eq:allocation-separator}
\end{align}
Such a choice is possible because
\begin{align}
    H_1(p)-H_1(q)
    \leq\Delta_R(p,q)<\rho_0.
\end{align}

The following lemma provides the
uniform spectral mass estimates
needed to verify the call inequalities.

\begin{lemma}[Tensor-power mass estimates]
\label{lem:tensor-mass}
For all sufficiently large integers $k$,
\begin{align}
    F_x(s)&\leq\frac14 F_y(Ts),
    \nonumber\\
    F_y(s)&\leq\frac14 F_y(Ts),
    \qquad s\leq2^{kc},
    \label{eq:allocation-lower-mass}
\end{align}
and
\begin{align}
    G_y(Ts)\leq\frac14 G_x(s),
    \qquad s\geq2^{kc}.
    \label{eq:allocation-upper-mass}
\end{align}
All inequalities hold simultaneously
for every threshold $s$ in the
indicated ranges.
\end{lemma}

\begin{proof}
We apply Lemma~\ref{lem:uniform-tail-comparison} with
\begin{align}
    A=Z_q-\rho_0,
    \qquad B=Z_p.
\end{align}
These variables have the same supports as those used for the count
inequalities, so they are nonconstant and satisfy the strict endpoint
comparisons required by the lemma.

For $\lambda>-1$,
$\lambda\neq0$, we have
\begin{align}
    &\log_2
    \frac{\mathbb E[2^{\lambda A}]}
         {\mathbb E[2^{\lambda B}]}=
    \lambda\left[
    H_{1+\lambda}(p)
    -H_{1+\lambda}(q)-\rho_0
    \right].
    \label{eq:mass-moment-comparison}
\end{align}

Since
$\Delta_R(p,q)<\rho_0$,
this expression is strictly
negative for $\lambda>0$
and strictly positive for
$-1<\lambda<0$.

At $\lambda=-1$,
the moment ratio is
$2^{\rho_0}>1$.
For $\lambda<-1$,
convexity of $u^{1+\lambda}$
and $p\prec q$ imply
\begin{align}
    \sum_iq_i^{1+\lambda}
    \geq\sum_ip_i^{1+\lambda},
\end{align}
so the moment ratio remains
strictly greater than one.

Moreover,
\begin{align}
    \mathbb E[A]
    =-H_1(q)-\rho_0
    <c<-H_1(p)
    =\mathbb E[B].
\end{align}

All hypotheses of
Lemma~\ref{lem:uniform-tail-comparison}
are therefore satisfied.

Using the probabilistic identities in
Eq.~\eqref{eq:spectral-mass-probabilities}, the lower-tail estimate
yields
\begin{align}
    F_x(s)\leq\frac14F_y(Ts),
    \qquad s\leq2^{kc},
\end{align}
while the upper-tail estimate gives
\begin{align}
    G_y(Ts)\leq\frac14G_x(s),
    \qquad s\geq2^{kc}.
\end{align}

It remains to prove the second
lower-tail inequality.

To this end, apply
Lemma~\ref{lem:uniform-tail-comparison}
to
\begin{align}
    A=Z_q-\rho_0,
    \qquad B=Z_q.
\end{align}
The moment ratio is now
\begin{align}
    \frac{\mathbb E[2^{\lambda A}]}
         {\mathbb E[2^{\lambda B}]}
    =2^{-\lambda\rho_0}.
\end{align}
Thus the strict moment inequalities
hold for every $\lambda\neq0$.
The mean and both support endpoints
are also strictly shifted downward
by $\rho_0$.

Furthermore, our choice of $c$
satisfies
\begin{align}
    -H_1(q)-\rho_0
    <c<-H_1(p)<-H_1(q),
\end{align}
so the separator lies strictly
between the two means.

The lower-tail estimate therefore
gives
\begin{align}
    F_y(s)\leq\frac14F_y(Ts),
    \qquad s\leq2^{kc}.
\end{align}

We can therefore choose $k$ sufficiently large
so that all three inequalities hold.
\end{proof}

We now use these estimates to prove the call inequalities
\eqref{eq:cum-call}.

Choose $k$ sufficiently large for
the estimates of
Lemma~\ref{lem:tensor-mass} to hold,
and such that
\begin{align}
    2T\leq\Gamma,
\end{align}
which is possible since $\rho_0<\rho$.

Fix $b>0$ and $0\leq a\leq b$.
We prove
\begin{align}
    J_x(a,b)\leq J_y(a,\Gamma b)
\end{align}
by considering two cases.

First, suppose that $2b\leq2^{kc}$.
Since $2T\leq\Gamma$ and $a\leq b$,
we have
\begin{align}
    J_y(a,\Gamma b)
    &\geq
    \sum_{2b<y_j\leq2Tb}(y_j-a)_+
    \nonumber\\
    &\geq
    \frac12\bigl[F_y(2Tb)-F_y(2b)\bigr]
    \nonumber\\
    &\geq
    \frac38 F_y(2Tb)
    \nonumber\\
    &\geq
    \frac32 F_x(2b)
    \nonumber\\
    &\geq J_x(a,b).
    \label{eq:allocation-truncated-low}
\end{align}
Here the second inequality follows from
$y_j>2b\geq2a$, while the third and
fourth follow from
Eq.~\eqref{eq:allocation-lower-mass}
at $s=2b$.
The last inequality uses
\begin{align}
    J_x(a,b)\leq F_x(b)\leq F_x(2b).
\end{align}

Next, suppose that $2b>2^{kc}$.
We compare the contributions excluded
from the full sums.

For the target spectrum,
Eq.~\eqref{eq:allocation-upper-mass}
at $s=2b$ gives
\begin{align}
    \sum_{y_j>\Gamma b}(y_j-a)_+
    &\leq\sum_{y_j>\Gamma b}y_j
    \nonumber\\
    &\leq G_y(2Tb)
    \nonumber\\
    &\leq\frac14G_x(2b).
    \label{eq:allocation-target-removed}
\end{align}
For the source spectrum, since $a\leq b$,
\begin{align}
    \sum_{x_i>b}(x_i-a)_+
    &\geq
    \sum_{x_i\geq2b}(x_i-a)_+
    \nonumber\\
    &\geq\frac12G_x(2b).
    \label{eq:allocation-source-removed}
\end{align}
Consequently,
\begin{align}
    \sum_{y_j>\Gamma b}(y_j-a)_+
    \leq
    \sum_{x_i>b}(x_i-a)_+.
\end{align}

On the other hand, since $x\prec y$
and $(u-a)_+$ is convex,
\begin{align}
    \sum_i(x_i-a)_+
    \leq
    \sum_j(y_j-a)_+.
\end{align}
Subtracting the contributions
excluded by the respective thresholds,
we conclude that
\begin{align}
    J_y(a,\Gamma b)
    &=
    \sum_j(y_j-a)_+
    -
    \sum_{y_j>\Gamma b}(y_j-a)_+
    \nonumber\\
    &\geq
    \sum_i(x_i-a)_+
    -
    \sum_{x_i>b}(x_i-a)_+
    \nonumber\\
    &=J_x(a,b).
    \label{eq:allocation-truncated-high}
\end{align}

Thus Eq.~\eqref{eq:cum-call} holds
in both cases, uniformly over
$b>0$ and $0\leq a\leq b$.

Together with the count inequalities
established above, this verifies
both conditions of
Lemma~\ref{lem:scalar-allocation}.

Hence Lemma~\ref{lem:cumulative}
provides a nonnegative matrix
$\tau=(\tau_{ij})$ satisfying
\begin{align}
    \sum_j\tau_{ij}=1,
    \qquad
    \sum_i\tau_{ij}=1,
    \qquad
    \sum_j\tau_{ij}y_j=x_i,
\end{align}
and
\begin{align}
    \tau_{ij}>0
    \quad\Longrightarrow\quad
    y_j\leq2^{k\rho}x_i.
\end{align}

For each source type $t$ and
target type $s$, define
\begin{align}
    n_{st}
    :=
    \sum_{i\text{ of type }t}
    \sum_{j\text{ of type }s}\tau_{ij}.
\end{align}
Since $x_i=z_t$ and $y_j=w_s$
for indices of types $t$ and $s$,
respectively, the properties of $\tau$
imply
\begin{align}
    \sum_t n_{st}&=N_s',
    &
    \sum_s n_{st}&=N_t,\\
    \sum_s w_s n_{st}&=N_tz_t.
\end{align}
Moreover,
\begin{align}
    n_{st}>0
    \quad\Longrightarrow\quad
    w_s\leq2^{k\rho}z_t.
\end{align}

These are precisely the conditions
of Lemma~\ref{lem:spectral-allocation},
completing its proof.

\section{Asymptotic communication rates}
\label{sec:rates}

We now turn to transformations involving multiple
copies of the source and target states. Rather than
implementing a one-copy protocol independently on
each copy, Alice and Bob may perform collective
local operations on their respective systems.
Such protocols may use a smaller message alphabet than
independent repetitions of an optimal one-copy
protocol.

For each positive integer $n$, let
\begin{align}
    B_n
    &:=
    B_{\rightarrow}
    (p^{\otimes n}\to q^{\otimes n}),
    \\
    C_n^*
    &:=
    C^*
    (p^{\otimes n}\to q^{\otimes n})
\end{align}
denote the minimum message alphabet sizes for
the collective transformation of $n$ copies,
without and with catalytic assistance,
respectively. In the latter case, the catalyst
may depend on $n$, but must remain finite
dimensional and be returned exactly and
uncorrelated with the target.

To characterize their
asymptotic growth, we consider the logarithm
of the alphabet size, normalized by the number
of copies. We therefore define the noncatalytic
and catalytic communication rates as
\begin{align}
    R_{\mathrm{nc}}(p,q)
    &:=
    \lim_{n\to\infty}
    \frac{\log_2 B_n}{n},
    \\
    R_{\mathrm{cat}}(p,q)
    &:=
    \lim_{n\to\infty}
    \frac{\log_2 C_n^*}{n}.
\end{align}
These quantities describe the asymptotic
number of classical bits per input copy,
under fixed-length encoding of the
communicated messages.

Both limits exist. Indeed, combining
protocols for $n$ and $k$ copies gives
\begin{align}
    B_{n+k}\leq B_nB_k,
    \qquad
    C_{n+k}^*\leq C_n^*C_k^*.
\end{align}
In the catalytic case, the two catalysts
can be combined into a single finite
catalyst. Hence both sequences are
submultiplicative, and their logarithms
are subadditive. By Fekete's lemma~\cite{Fekete1923},
the normalized sequences $\log_2 B_n/n$ and $\log_2 C_n^*/n$
therefore converge to finite limits.

We show that, despite the potentially unbounded compression achievable
for individual conversions, catalytic assistance does not improve
the asymptotic communication rate. The following theorem establishes
that the rates with and without a catalyst coincide and are both
given by the R\'enyi gap $\Delta_R(p,q)$.

\begin{theorem}[Asymptotic communication rates]
\label{thm:asymptotic-rates}
Let $p,q$ be distinct, full-support probability vectors of equal length satisfying $p\prec q$. Then
\begin{align}
    R_{\mathrm{nc}}(p,q)
    =
    R_{\mathrm{cat}}(p,q)
    =
    \Delta_R(p,q).
    \label{eq:asymptotic-rates}
\end{align}
\end{theorem}

The catalytic rate follows directly from
Theorem~\ref{thm:catalytic-cost} and the additivity of R\'enyi entropies.
Indeed, the exact catalytic cost at every blocklength is
\begin{align}
    C_n^*
    =
    \left\lfloor 2^{n\Delta_R(p,q)}\right\rfloor+1.
    \label{eq:multicopy-catalytic-cost}
\end{align}
Since $C_n^*\leq B_n$, this also provides the claimed lower bound on
the noncatalytic rate. To obtain the matching upper bound, we exploit
exact catalyst return: a fixed catalyst can be prepared, reused across
many blocks, and removed with an additional communication cost
independent of the number of blocks.
This fixed overhead is spread over all input copies, so its
contribution to the communication rate vanishes as the number
of copies grows. We refer to this mechanism as \emph{amortization}.

\subsection{Exact amortization}
\label{subsec:amortization}

The following proposition makes this amortization argument precise.
The argument applies to arbitrary finite catalyst rank and does not
require the source and target Schmidt vectors to have distinct extreme entries.
\begin{proposition}[Exact amortization]
\label{prop:amortization}
Suppose that a finite pure catalyst $r$
allows
\begin{align}
    \Bto\left(
    p^{\otimes b}\otimes r
    \to q^{\otimes b}\otimes r
    \right)\leq m
\end{align}
for some integers $b,m\geq1$.

Then there exists a finite constant
$A$, independent of $n$, such that
\begin{align}
    B_n\leq A\,m^{\lfloor n/b\rfloor}
    \label{eq:amortization}
\end{align}
for every $n\geq0$, with $B_0=1$.
\end{proposition}

We use two technical lemmas, proved in
Secs.~\ref{app:fixed-offset} and~\ref{app:uniform-cancel}
of Appendix~\ref{app:amortization}, respectively.

\begin{lemma}[Absorption of a fixed auxiliary spectrum]
\label{lem:offset}
Let $a,b$ be strictly positive
probability vectors of equal length,
satisfying
\begin{align}
    a\prec b,\qquad
    a_{\max}<b_{\max},\qquad
    a_{\min}>b_{\min}.
\end{align}
For every strictly positive probability
vector $r$ of length $h$, there exists
an integer $K\geq1$ such that
\begin{align}
    r\otimes a^{\otimes K}
    \prec u_h\otimes b^{\otimes K}.
    \label{eq:fixed-offset}
\end{align}
\end{lemma}

\begin{lemma}[Uniform-factor cancellation]
\label{lem:uniform-cancel}
For strictly positive probability vectors
$x,y$ of equal length and any $h\geq1$,
\begin{align}
    \Bto(x\otimes u_h\to y\otimes u_h)
    =
    \Bto(x\to y).
    \label{eq:uniform-cancellation}
\end{align}
\end{lemma}

We now prove Proposition~\ref{prop:amortization}.

\begin{proof}[Proof of Proposition~\ref{prop:amortization}]
Let $h$ be the Schmidt rank of $r$.

\emph{Step 1: Strict core.}
Let $i_0$ and $i_1$ be the first and last
indices where $p$ and $q$ differ.
Since $p\prec q$,
\begin{align}
    p_{i_0}<q_{i_0},
    \qquad
    p_{i_1}>q_{i_1}.
\end{align}
The corresponding intervals have equal
total mass $t>0$. Normalizing these
intervals and collecting the remaining
common entries yields
\begin{align}
    p=(1-t)c\oplus tp_0,
    \qquad
    q=(1-t)c\oplus tq_0.
    \label{eq:strict-core}
\end{align}
Here $p_0,q_0$ have dimension
$d_0=i_1-i_0+1\geq2$ and satisfy
\begin{align}
    p_0\prec q_0,
    \qquad
    (p_0)_{\max}<(q_0)_{\max},
    \qquad
    (p_0)_{\min}>(q_0)_{\min}.
\end{align}
If $t=1$, the common block is absent.

\emph{Step 2: Catalyst buffer.}
By Lemma~\ref{lem:offset}, there exists
an integer $K\geq1$ such that
\begin{align}
    r\otimes p_0^{\otimes K}
    \prec
    u_h\otimes q_0^{\otimes K}.
\end{align}
Combining this with $u_h\prec r$ gives
\begin{align}
\begin{aligned}
    u_h\otimes p_0^{\otimes K}
    &\prec r\otimes p_0^{\otimes K}\\
    &\prec u_h\otimes q_0^{\otimes K}.
\end{aligned}
    \label{eq:buffer-chain}
\end{align}

\emph{Step 3: Amortization with reserved copies.}
Write
\begin{align}
    n=Lb+s,
    \qquad
    L=\left\lfloor\frac nb\right\rfloor,
    \qquad 0\leq s<b.
\end{align}

Consider the following sequence of
exact transformations:

\begin{enumerate}
    \item Prepare $r$ from $u_h$,
    leaving $p_0^{\otimes K}$ unchanged:
    a message alphabet of size at most $h$.

    \item Apply the catalytic protocol
    to the $L$ blocks of $b$ copies:
    communication alphabet size at most $m^L$.

    \item Convert the remaining
    $s$ copies without assistance:
    a message alphabet of size at most $d^s\leq d^{b-1}$.

    \item Convert
    $r\otimes p_0^{\otimes K}$
    into
    $u_h\otimes q_0^{\otimes K}$:
    a message alphabet of size at most $hd_0^K$.
\end{enumerate}

The dimension bounds in steps 1, 3,
and 4 follow from majorization and
Carath\'eodory's theorem.
Exact catalyst return ensures that
the same $r$ is available in every
block of step 2.

Thus the complete protocol has communication
alphabet size at most
\begin{align}
    A\,m^L,
    \qquad
    A:=h^2d_0^Kd^{b-1},
    \label{eq:amortization-constant}
\end{align}
where $A$ is independent of $n$.

The uniform auxiliary state $u_h$
is restored exactly. Hence
Lemma~\ref{lem:uniform-cancel} gives
\begin{align}
\begin{aligned}
    &\Bto\left(
    p_0^{\otimes K}\otimes p^{\otimes n}
    \to
    q_0^{\otimes K}\otimes q^{\otimes n}
    \right)\\
    &\qquad\leq
    A\,m^{\lfloor n/b\rfloor}.
\end{aligned}
    \label{eq:reserved-copy-bound}
\end{align}

\emph{Step 4: Elimination of reserved copies.}

The preceding bound still involves
$K$ additional copies of $p_0$.
We now show that they can be removed
without increasing the communication bound.

For $0\leq j\leq K$, define
\begin{align}
    F_j(n):=
    \Bto\left(
    p_0^{\otimes j}\otimes p^{\otimes n}
    \to
    q_0^{\otimes j}\otimes q^{\otimes n}
    \right),
\end{align}
and write
\begin{align}
    M_n:=A\,m^{\lfloor n/b\rfloor}.
\end{align}

We aim to prove
\begin{align}
    F_j(n)\leq M_n
    \qquad
    (0\leq j\leq K,\ n\geq0).
    \label{eq:amortization-induction-goal}
\end{align}

We already know that
\begin{align}
    F_K(n)\leq M_n
    \qquad(n\geq0)
\end{align}
from Eq.~\eqref{eq:reserved-copy-bound}.

For $j<K$ and $n\geq1$, we use
Eq.~\eqref{eq:strict-core} to decompose
one of the $n$ copies.

Up to permutations of coordinates,
the source vector
$p_0^{\otimes j}\otimes p^{\otimes n}$
becomes
\begin{align}
    &(1-t)
    \left(c\otimes p_0^{\otimes j}
    \otimes p^{\otimes(n-1)}\right)
    \oplus
    t\left(p_0^{\otimes(j+1)}
    \otimes p^{\otimes(n-1)}\right),
\end{align}
with an analogous decomposition
of the target.

In the first block, $c$ is unchanged, and the remaining
transformation admits a message alphabet of size at most
$F_j(n-1)$. The second block admits a message alphabet
of size at most $F_{j+1}(n-1)$.

Since the corresponding block weights
are identical, the direct-sum rule
in Eq.~\eqref{eq:direct-sum-budget}
gives
\begin{align}
    F_j(n)\leq
    \max\{F_j(n-1),F_{j+1}(n-1)\}.
    \label{eq:amortization-recursion}
\end{align}
When $t=1$, only the second block
is present.

We now prove
Eq.~\eqref{eq:amortization-induction-goal}
by induction on $n$.

For $n=0$, the majorization
$p_0\prec q_0$ implies
$p_0^{\otimes j}\prec q_0^{\otimes j}$.
Since these vectors have dimension
$d_0^j$, the dimension bound gives
\begin{align}
    F_j(0)\leq d_0^j
    \leq d_0^K\leq A=M_0,
    \qquad 0\leq j\leq K.
\end{align}

Suppose the claim holds for $n-1$,
simultaneously for all $j$.

For $j=K$, the desired inequality
is already established.

For $j<K$, the recurrence and
the induction hypothesis imply
\begin{align}
    F_j(n)
    &\leq
    \max\{F_j(n-1),F_{j+1}(n-1)\}
    \nonumber\\
    &\leq M_{n-1}
    \leq M_n,
\end{align}
where the last inequality follows
from $m\geq1$.

Thus the claim holds for every
$j$ and $n$. In particular,
taking $j=0$ gives
\begin{align}
    B_n=F_0(n)
    \leq A\,m^{\lfloor n/b\rfloor},
\end{align}
as required.
\end{proof}

\subsection{Proof of the asymptotic rate theorem}

\begin{proof}[Proof of Theorem~\ref{thm:asymptotic-rates}]
Let $\Delta=\Delta_R(p,q)$.
Equation~\eqref{eq:multicopy-catalytic-cost} gives
\begin{align}
    2^{n\Delta}<C_n^*\leq 2^{n\Delta+1},
\end{align}
which implies
\begin{align}
    R_{\mathrm{cat}}(p,q)=\Delta.
\end{align}
Since $B_n\geq C_n^*$, we also have
\begin{align}
    \liminf_{n\to\infty}\frac{\log_2 B_n}{n}\geq\Delta.
\end{align}

To establish the reverse inequality, fix an arbitrary integer $b\geq1$. Theorem~\ref{thm:catalytic-cost}, applied to $b$ copies,
guarantees the existence of a finite pure catalyst
permitting the transformation with a message alphabet
of size
\begin{align}
    m_b:=\lfloor 2^{b\Delta}\rfloor+1
    \leq 2^{b\Delta+1}.
\end{align}

By Proposition~\ref{prop:amortization}, there exists a finite constant $A_b$, independent of $n$, such that
\begin{align}
    B_n\leq A_b\,m_b^{\lfloor n/b\rfloor}.
\end{align}
It follows that
\begin{align}
    B_n
    \leq A_b\,2^{(b\Delta+1)\lfloor n/b\rfloor}
    \leq A_b\,2^{n(\Delta+1/b)}.
\end{align}

Combining the upper and lower bounds, we obtain, for every $n\geq1$,
\begin{align}
    \Delta
    \leq \frac{\log_2 B_n}{n}
    \leq \Delta+\frac1b+\frac{\log_2 A_b}{n}.
    \label{eq:rate-squeeze}
\end{align}

Taking $n\to\infty$ with $b$ fixed gives
\begin{align}
    \Delta
    \leq \liminf_{n\to\infty}\frac{\log_2 B_n}{n}
    \leq \limsup_{n\to\infty}\frac{\log_2 B_n}{n}
    \leq \Delta+\frac1b.
\end{align}
Since $b$ is arbitrary, letting $b\to\infty$ yields
\begin{align}
    R_{\mathrm{nc}}(p,q)=\Delta.
\end{align}

Together with the catalytic rate, this proves
\begin{align}
    R_{\mathrm{nc}}(p,q)
    =
    R_{\mathrm{cat}}(p,q)
    =
    \Delta_R(p,q).
\end{align}
The catalyst and the constant $A_b$ may depend on $b$, but their finiteness for each fixed $b$ is sufficient.
\end{proof}

As a consequence, for any fixed pair $p,q$ satisfying our assumptions,
\begin{align}
    \frac{B_n}{C_n^*}=2^{o(n)}.
\end{align}
Thus catalysis cannot provide an exponential communication advantage in the number of copies, although subexponential savings are not excluded.

\section{Conclusion}
\label{sec:discussion}

In this work, we have studied the classical communication
required for exact deterministic transformations of
bipartite pure states in the presence of an entanglement
catalyst. For distinct full-support Schmidt vectors
of equal length satisfying $p\prec q$, we have obtained
an exact expression for the minimum catalytic message
alphabet size in terms of the maximum R\'enyi entropy
difference. We have also shown that this minimum can
always be achieved with a finite pure catalyst returned
exactly and uncorrelated with the target.

The significance of this result is that convertibility
alone does not capture the usefulness of a catalyst.
Even when catalytic assistance is unnecessary for making
a transformation possible, it can reduce the classical
communication needed to implement it, without consuming
the auxiliary entanglement. Our formula makes this role
quantitative by determining the optimal catalytic
alphabet size directly from the source and target spectra.

We have studied, in particular, the family constructed
in Sec.~\ref{sec:maximal-compression}, for which the
minimum unassisted alphabet size is $d$, the largest
possible for Schmidt vectors of length $d$, while the
minimum catalytic alphabet size is two. The reduction
factor therefore grows without bound with the Schmidt
rank. We have then considered the limit of many copies
of a fixed source and target. In this case, we have
proved that the catalytic and noncatalytic communication
rates coincide and are given by the same R\'enyi entropy
difference that determines the exact catalytic cost.
Thus the communication saving for individual
transformations does not persist at the level
of the leading asymptotic rate.

Our asymptotic result complements studies of
catalytic entanglement distillation and concentration
\cite{GanardiKondraStreltsov2024,Streltsov2026SecondOrder}.
These works concern the entanglement yield,
while we have investigated the classical
communication required for exact state transformations.

The assumption of equal Schmidt rank plays
an important role in our analysis.
It would be interesting to extend these results
to transformations between states with different
Schmidt ranks. Further directions would be
to consider mixed catalysts or relax the
conditions on how the catalyst is returned.

It would also be interesting to obtain a similarly
compact expression for the minimum noncatalytic
message alphabet size $\Bto(p\to q)$, which our
analysis has not provided. Such an expression
would allow a more direct evaluation of the
communication saving offered by a catalyst
for a given source and target.

Finally, the equality of the asymptotic rates
does not exclude a catalytic advantage that
grows with the number of copies.
For example, a polynomially growing ratio
between the noncatalytic and catalytic
alphabet sizes would be compatible with
equal rates. It would be interesting to
determine whether such growth can occur
for a fixed source and target, and how
large the resulting advantage can be.

\section*{Acknowledgments}
We thank Jofre Abellanet Vidal for discussion. 
\section*{Use of AI tools}

During the preparation of this manuscript, the authors
used ChatGPT (OpenAI) and Claude (Anthropic) to assist
with language editing, literature searches, and
discussions of mathematical arguments.
The authors take full responsibility for the accuracy
and integrity of all derivations, results, references,
and conclusions presented in this work.

\appendices

\section{Operational characterization}
\label{app:operational}

We establish the correspondence between
the message alphabet size in an
exact one-way LOCC transformation and
the number of terms in a convex
decomposition over a unitary orbit.
Explicit constructions using a single local measurement
and permutation corrections were given by Jensen and
Schack~\cite{JensenSchack2001}.

The following result first characterizes
such transformations using arbitrary
convex weights. The equal-weight
formulation employed throughout the
main text then follows from the
equalization result of
Daftuar and Hayden~\cite{DaftuarHayden2005}.

\begin{proposition}[Message alphabets and unitary decompositions]
\label{prop:operational-characterization}
Let $x,y$ be strictly positive
probability vectors of equal length,
and let
\begin{align}
    X=\operatorname{diag}(x),
    \qquad
    Y=\operatorname{diag}(y).
\end{align}
An exact one-way LOCC transformation
$x\to y$ can be implemented with
a message alphabet of size at most $m$ if
and only if there exist probabilities
$w_1,\ldots,w_m$ and unitaries
$U_1,\ldots,U_m$ such that
\begin{align}
    X=\sum_{j=1}^{m}
    w_jU_jYU_j^\dagger,
    \qquad
    w_j\geq0,
    \qquad
    \sum_{j=1}^{m}w_j=1.
    \label{eq:operational-unitary-average}
\end{align}

Equivalently, by
Corollary~2.7 of
Daftuar and Hayden~\cite{DaftuarHayden2005},
\begin{align}
    B_{\to}(x\to y)\leq m
    \quad\Longleftrightarrow\quad
    X\in\mathcal F_m(Y).
    \label{eq:operational-equal-weight}
\end{align}
\end{proposition}

\begin{proof}
We first show that the unitary decomposition
\eqref{eq:operational-unitary-average}
gives rise to an exact one-way LOCC protocol
with a message alphabet of size at most $m$.

Suppose that
\begin{align}
    X=\sum_{j=1}^{m}w_jU_jYU_j^\dagger,
    \qquad
    w_j\geq0,
    \qquad
    \sum_{j=1}^{m}w_j=1.
\end{align}
Alice performs a measurement with Kraus
operators
\begin{align}
    M_j
    :=
    \sqrt{w_j}\,
    Y^{1/2}U_j^\dagger X^{-1/2}.
    \label{eq:operational-Alice-measurement}
\end{align}
Since $X$ is positive definite, these
operators are well defined. Moreover,
\begin{align}
    \sum_{j=1}^{m}M_j^\dagger M_j
    &=
    X^{-1/2}
    \left(
    \sum_{j=1}^{m}w_jU_jYU_j^\dagger
    \right)
    X^{-1/2}
    \nonumber\\
    &=I.
\end{align}
Thus they define a valid quantum
measurement.

Upon obtaining outcome $j$, Alice
communicates the label to Bob, who
applies the unitary $U_j^{\mathsf T}$.

To verify the resulting transformation,
recall that a bipartite pure state can
be identified with its coefficient
matrix $C$, under which local
operators $A\otimes B$ act as
\begin{align}
    C\longmapsto ACB^{\mathsf T}.
\end{align}
The source state has coefficient
matrix $X^{1/2}$. Therefore, in
branch $j$, the coefficient matrix
after Alice's measurement and
Bob's correction is
\begin{align}
    M_jX^{1/2}(U_j^{\mathsf T})^{\mathsf T}
    &=
    \sqrt{w_j}\,
    Y^{1/2}U_j^\dagger U_j
    \nonumber\\
    &=\sqrt{w_j}\,Y^{1/2}.
    \label{eq:operational-branch-matrix}
\end{align}
Hence outcome $j$ occurs with
probability $w_j$, and the normalized
output is exactly the target state
with Schmidt coefficients $\sqrt{y}$.

The protocol therefore implements
$x\to y$ exactly, using a message alphabet of size at most $m$.

Conversely, suppose that an exact
one-way LOCC protocol transforms
$x$ into $y$ using a message alphabet of size at most $m$.

Let $\{\mathcal A_j\}_{j=1}^m$
denote Alice's quantum instrument,
where $j$ labels the communicated
message. Each $\mathcal A_j$ is
completely positive and trace
nonincreasing, and
$\sum_j\mathcal A_j$ is trace preserving.
The instrument may contain several
Kraus operators associated with the
same message, corresponding to local
outcomes that are not communicated.

Conditional on message $j$, Bob
applies a trace-preserving quantum
channel $\mathcal B_j$.
By introducing local ancillas,
we may represent Alice's conditional
operation and Bob's channel by
Stinespring dilations, using a contraction
for Alice and an isometry for Bob,
while retaining all uncommunicated
outcomes coherently.

For each message $j$ occurring with
nonzero probability $w_j$, the
conditional global state is therefore
pure. Since the required output is
the pure target state $|\phi_y\rangle$,
the final state, including the local
environments, must have the form
\begin{align}
    |\phi_y\rangle_{AB}
    \otimes|\eta_j\rangle_{A'B'}.
\end{align}

Local linear operations cannot increase
the Schmidt rank of a pure state.
Since the initial and target states
both have Schmidt rank $d$, we obtain
\begin{align}
    d\,\operatorname{rank}_{\mathrm{Sch}}
    (|\eta_j\rangle)\leq d.
\end{align}
Hence $|\eta_j\rangle$ is a product
state across Alice's and Bob's
environments.

It follows that Bob's reduced state,
immediately before his conditional
channel, has the same nonzero
eigenvalues as $Y$. Indeed, Bob's
Stinespring isometry preserves the
nonzero eigenvalues of his reduced
state, while the final environment
on his side is pure.

Denoting this conditional state by
$\rho_B^{(j)}$, we therefore have
\begin{align}
    \rho_B^{(j)}
    =U_jYU_j^\dagger
\end{align}
for some unitary $U_j$.

Finally, Alice's instrument is trace
preserving when averaged over its
outcomes and therefore leaves Bob's
reduced state unchanged on average.
Consequently,
\begin{align}
    X=\sum_{j=1}^{m}w_j\rho_B^{(j)}
    =\sum_{j=1}^{m}
    w_jU_jYU_j^\dagger,
\end{align}
where $w_j\geq0$ and
$\sum_jw_j=1$.

This establishes the characterization
in terms of arbitrary convex weights.
By Corollary~2.7 of
Daftuar and Hayden~\cite{DaftuarHayden2005},
the decomposition can be replaced
by an equal-weight average of $m$
unitary conjugates, proving
Eq.~\eqref{eq:operational-equal-weight}.
\end{proof}

\section{Reduction to one-way LOCC}
\label{app:oneway}

We now show that, for the pure-state transformations
considered here, allowing multiple rounds of classical
communication does not reduce the minimum
communication alphabet size.

For a finite-round LOCC protocol, the communication
alphabet consists of the complete sequences of
exchanged classical messages that occur with nonzero
probability. Each sequence records the communicated
messages in their order, from the start of the protocol
to its termination. Uncommunicated local outcomes
are not included.

Our argument is related to the one-way reductions
of Lo and Popescu~\cite{LoPopescu2001} and Harrow
and Lo~\cite{HarrowLo2004}, but explicitly preserves
both the alphabet size and the associated probabilities.
By Proposition~\ref{prop:operational-characterization},
it suffices to establish a convex unitary decomposition
with one term for each complete sequence.

\begin{proposition}[One-way reduction]
\label{prop:oneway}
Let $p,q$ be strictly positive probability vectors of equal length, and let $\Pi$ be an exact deterministic finite-round LOCC protocol implementing $p\to q$. Suppose that its communication alphabet consists of $N$ complete sequences, occurring with probabilities
$w_1,\ldots,w_N>0$. Then there exist unitaries $U_1,\ldots,U_N$ such that
\begin{align}
    P=\sum_{\ell=1}^{N}w_\ell U_\ell Q U_\ell^\dagger,
    \label{eq:oneway-leaf-average}
\end{align}
where $P=\operatorname{diag}(p)$ and $Q=\operatorname{diag}(q)$. Consequently, the transformation admits an exact deterministic one-way LOCC protocol with the same message probabilities.
\end{proposition}

\begin{proof}
Let $|\psi_p\rangle$ and $|\phi_q\rangle$
be the initial and target bipartite pure
states, with squared Schmidt coefficients
$p$ and $q$, respectively.

Consider an exact deterministic finite-round
LOCC protocol $\Pi$ transforming
$|\psi_p\rangle$ into $|\phi_q\rangle$.
We represent the protocol by a finite rooted
tree, whose nodes correspond to partial
sequences of exchanged classical messages
and whose leaves correspond to complete sequences.

The root is labeled by the empty sequence
$\varnothing$. At each internal node $s$,
one of the parties performs a local
instrument and communicates an outcome $j$,
leading to the child node $sj$ with
conditional probability $t_{j|s}$.
We denote by $N$ the number of leaves,
corresponding to the possible complete
sequences of exchanged classical messages
of the protocol.

By coherently retaining uncommunicated
local outcomes in ancillary systems,
we may assume that the joint state
$|\xi_s\rangle$, conditioned on each
sequence $s$ of exchanged classical messages,
is pure.

Let $d$ be the common Schmidt rank of the initial and target states. At the end of any complete sequence $\ell$ of
exchanged classical messages, exactness and purity
of the target imply that the conditional global
state has the form
\begin{align}
    |\xi_\ell\rangle=|\phi_q\rangle\otimes|\eta_\ell\rangle,
\end{align}
where $|\eta_\ell\rangle$ describes the ancillary systems. Since local operations cannot increase Schmidt rank,
\begin{align}
    d\,\operatorname{rank}_{\mathrm{Sch}}(\eta_\ell)\leq d.
\end{align}
Consequently, $|\eta_\ell\rangle$ is a product state, and the squared Schmidt spectrum of $|\xi_\ell\rangle$ is $q$. Moreover, every intermediate conditional state must also have Schmidt rank $d$, since its rank cannot exceed that of the initial state or be smaller than that of any possible final state.

For each sequence $s$, let $D_s$ be the diagonal matrix of the squared Schmidt coefficients of the corresponding conditional state. In particular,
\begin{align}
    D_{\varnothing}=P,\qquad D_\ell=Q
\end{align}
for every complete sequence $\ell$.

Consider now any intermediate sequence $s$. Suppose that Alice performs the next local instrument and communicates the outcome $j$, which occurs with conditional probability $t_{j|s}$. Since Alice acts locally, Bob's reduced state is unchanged when averaged over the possible outcomes:
\begin{align}
    \rho_s=\sum_j t_{j|s}\rho_{sj}.
\end{align}
Here $\rho_s$ and $\rho_{sj}$ denote Bob's reduced states before and after the corresponding outcome.

All these matrices have rank $d$, and positivity of the decomposition implies that their supports coincide. Since the global conditional states are pure, their nonzero spectra are precisely those of $D_s$ and $D_{sj}$. Hence, upon diagonalization on the common support, there exist unitaries $U_{s,j}$ such that
\begin{align}
    D_s=\sum_j t_{j|s}U_{s,j}D_{sj}U_{s,j}^\dagger.
    \label{eq:oneway-node}
\end{align}
The same identity holds when Bob performs the local instrument, by considering Alice's reduced state instead.

We can now apply this identity successively to the conditional states appearing on its right-hand side, until every sequence is complete. Since the squared Schmidt spectrum of every final state is $q$, we obtain
\begin{align}
    P=\sum_{\ell=1}^{N}w_\ell V_\ell QV_\ell^\dagger,
    \label{eq:oneway-leaf-average-proof}
\end{align}
where $w_\ell$ is the probability of the complete
sequence $\ell$, obtained by multiplying the
conditional probabilities of its successive messages,
and $V_\ell$ is the corresponding product of unitaries.

By Proposition~\ref{prop:operational-characterization}, this decomposition yields an exact one-way LOCC protocol with a message alphabet of size $N$ and the same probabilities as the original protocol.
\end{proof}

The proposition implies that the minimum communication
alphabet size over finite-round LOCC protocols
coincides with the minimum one-way message alphabet
size $B_{\rightarrow}(p\to q)$. The same conclusion holds for catalytic transformations by applying the proposition to the enlarged source and target Schmidt vectors $p\otimes r$ and $q\otimes r$. In particular, optimizing over finite pure catalysts does not change this equivalence.

\section{Uniform-source construction}
\label{app:uniform}

In this appendix, we establish the exact
multicopy decomposition used in the proof
of Proposition~\ref{prop:uniform-source}.

Our argument relies on the additive Horn
theorem, which characterizes the possible
spectra of sums of Hermitian matrices.
We first express the required decomposition
in terms of Horn inequalities. We then
derive three spectral estimates for tensor
powers and use them to verify all the
inequalities simultaneously.

\subsection{Reduction to Horn inequalities}
\label{app:uniform-horn}

We recall the additive Horn theorem
\cite{Klyachko1998,KnutsonTao1999}
in its Littlewood--Richardson formulation,
as stated in Ref.~\cite[Theorem~17]{Fulton2000}.

\begin{theorem}[Additive Horn theorem]
\label{fact:horn}
Let
$\alpha^{(1)},\ldots,\alpha^{(M)},\gamma$
be nonincreasing real sequences of
length $D$ satisfying
\begin{align}
    \sum_i\gamma_i
    =
    \sum_{j=1}^{M}\sum_i\alpha_i^{(j)}.
\end{align}
There exist Hermitian matrices
$A_1,\ldots,A_M$ with spectra
$\alpha^{(1)},\ldots,\alpha^{(M)}$,
respectively, whose sum has spectrum
$\gamma$, if and only if
\begin{align}
    \sum_{i\in K}\gamma_i
    \leq
    \sum_{j=1}^{M}\sum_{i\in I_j}
    \alpha_i^{(j)}
    \label{eq:horn-general}
\end{align}
for every $1\leq r<D$ and every
tuple $(I_1,\ldots,I_M,K)$ of
$r$-element index sets with positive
Littlewood--Richardson coefficient
\begin{align}
    c_{\lambda(I_1),\ldots,
    \lambda(I_M)}^{\lambda(K)}>0.
\end{align}
Here, for
$I=\{i_1<\cdots<i_r\}$, we define
\begin{align}
    \lambda(I)
    =(i_r-r,\ldots,i_1-1).
\end{align}
\end{theorem}

We apply this characterization to
the following exact multicopy problem.

\begin{proposition}[Exact multicopy decomposition
from a uniform source]
\label{prop:uniformcase}
Let $q=(q_1,\ldots,q_d)\neq u_d$
be a full-support probability vector
in nonincreasing order, and let
$m\geq2$ be an integer.

If $dq_1<m$, then, for every
sufficiently large integer $k$,
there exist unitaries
$U_1,\ldots,U_{m^k}$ such that
\begin{align}
    \frac{I_{d^k}}{d^k}
    =
    \frac1{m^k}\sum_{j=1}^{m^k}
    U_j Q^{\otimes k}U_j^\dagger,
    \qquad Q=\operatorname{diag}(q).
    \label{eq:uniform-tensor-average}
\end{align}
In particular,
\begin{align}
    B_{\to}(u_{d^k}\to q^{\otimes k})
    \leq m^k
\end{align}
without a catalyst.
\end{proposition}

To prove the proposition, set
\begin{align}
    N=d^k,\qquad M=m^k,
    \qquad y=(q^{\otimes k})^\downarrow,
\end{align}
where $y_1\geq\cdots\geq y_N>0$.

By Theorem~\ref{fact:horn}, the desired
decomposition exists if and only if
the corresponding Horn inequalities
hold for $M$ summands, each with
spectrum $y/M$, and a sum with
spectrum $u_N$.

The trace condition is automatic.
Moreover, since the prescribed sum
has a uniform spectrum, each
admissible Horn inequality reduces to
\begin{align}
    \frac1M\sum_{j=1}^{M}
    \sum_{i\in I_j}y_i
    \geq\frac{\ell}{N},
    \label{eq:uniform-horn-condition}
\end{align}
where $1\leq\ell<N$ is the common
cardinality of the index sets.

The difficulty is therefore to verify
Eq.~\eqref{eq:uniform-horn-condition}
for every admissible Horn tuple.
We first establish the spectral
estimates needed for this purpose.

\subsection{Spectral estimates for tensor powers}
\label{app:uniform-tails}

Define two spectral cutoffs by
\begin{align}
    h_k:&=\#\left\{i:y_i\geq\frac2N\right\},\\
    \qquad
    g_k:&=\#\left\{i:y_i\geq\frac1{2N}\right\}.
    \label{eq:uniform-cutoffs}
\end{align}
Since $y$ is nonincreasing,
these count the eigenvalues above
twice and half the uniform eigenvalue,
respectively. In particular,
$h_k\leq g_k$.

We also define
\begin{align}
    S_k(t):=\sum_{i=1}^{t}y_i,
    \qquad
    \delta_k:=\frac{Ny_1}{M}.
\end{align}

The following three estimates will
allow us to control the Horn inequalities.

\begin{lemma}[Uniform spectral estimates]
\label{lem:uniform-spectral-estimates}
Under the assumptions of
Proposition~\ref{prop:uniformcase},
for every sufficiently large integer
$k$, the following inequalities hold
simultaneously:
\begin{subequations}
\label{eq:uniform-three-conditions}
\begin{align}
    h_k&\geq\frac{4N}{M},
    \label{eq:uniform-condition-h}\\
    \delta_k&\leq\frac14,
    \label{eq:uniform-condition-delta}\\
    S_k\left(\left\lfloor
    \frac{h_k}{4}\right\rfloor\right)
    &\geq
    2\delta_k+\frac{2g_k}{N}.
    \label{eq:uniform-condition-mass}
\end{align}
\end{subequations}
\end{lemma}

\begin{proof}
Since $y_1=q_1^k$, we have
\begin{align}
    \delta_k
    =\frac{Ny_1}{M}
    =\left(\frac{dq_1}{m}\right)^k
    \longrightarrow0,
    \label{eq:uniform-delta-limit}
\end{align}
where the limit follows from
$dq_1<m$.
This proves
Eq.~\eqref{eq:uniform-condition-delta}
for all sufficiently large $k$.

To establish the remaining estimates,
we use the weak law of large numbers
under two different sampling laws
\cite{CoverThomas2006}.

First, let $I_1,\ldots,I_k$ be
independent random indices with
$\Pr(I_a=i)=q_i$.
The probability of a word
$(i_1,\ldots,i_k)$ is then precisely
the corresponding eigenvalue
$q_{i_1}\cdots q_{i_k}$.

Consequently,
\begin{align}
\begin{split}
    S_k(h_k)
    =\Pr\Biggl(
    \frac1k\sum_{a=1}^{k}
    -\log_2q_{I_a}
    \leq\log_2d-\frac1k
    \Biggr).
\end{split}
    \label{eq:uniform-mass-probability}
\end{align}
The sample mean converges in
probability to $H_1(q)$.
Since $q$ is nonuniform,
\begin{align}
    H_1(q)<\log_2d,
\end{align}
and the weak law gives
\begin{align}
    S_k(h_k)\longrightarrow1.
    \label{eq:uniform-prefix-mass-limit}
\end{align}

Next, let $J_1,\ldots,J_k$
be independent indices uniformly
distributed over $\{1,\ldots,d\}$.
Every word then has probability
$1/N$, and
\begin{align}
\begin{split}
    \frac{g_k}{N}
    =\Pr\Biggl(
    \frac1k\sum_{a=1}^{k}
    -\log_2q_{J_a}
    \leq\log_2d+\frac1k
    \Biggr).
\end{split}
    \label{eq:uniform-count-probability}
\end{align}
The corresponding mean is
\begin{align}
    -\frac1d\sum_{i=1}^{d}\log_2q_i
    >\log_2d,
\end{align}
where the strict inequality follows
from Jensen's inequality and the
nonuniformity of $q$.
Therefore,
\begin{align}
    \frac{g_k}{N}\longrightarrow0.
    \label{eq:uniform-count-fraction-limit}
\end{align}

We now use these limits to establish
the two remaining estimates.

Since the first $h_k$ eigenvalues
are bounded above by $y_1=q_1^k$,
\begin{align}
    h_kq_1^k\geq S_k(h_k).
\end{align}
It follows that
\begin{align}
    \frac{h_k}{N/M}
    \geq
    S_k(h_k)
    \left(\frac{m}{dq_1}\right)^k
    \longrightarrow\infty.
    \label{eq:uniform-count-growth}
\end{align}
Hence
$h_k\geq4N/M$
for sufficiently large $k$,
proving
Eq.~\eqref{eq:uniform-condition-h}.

Furthermore, since $q_1<1$
and $S_k(h_k)\to1$, we also have
$h_k\to\infty$.

Because $y$ is nonincreasing,
the mean of its first
$\lfloor h_k/4\rfloor$ entries
is at least the mean of its first
$h_k$ entries. Therefore,
\begin{align}
    S_k\left(\left\lfloor
    \frac{h_k}{4}\right\rfloor\right)
    \geq
    \frac{\lfloor h_k/4\rfloor}{h_k}
    S_k(h_k).
    \label{eq:uniform-quarter-mass}
\end{align}
The right-hand side tends to $1/4$,
whereas
\begin{align}
    2\delta_k+\frac{2g_k}{N}
    \longrightarrow0.
\end{align}
Thus
Eq.~\eqref{eq:uniform-condition-mass}
also holds for sufficiently large $k$.

A single finite threshold on $k$
therefore guarantees all three
inequalities simultaneously.
\end{proof}

\subsection{Verification of the Horn inequalities}
\label{app:uniform-verification}

We can now complete the proof of
Proposition~\ref{prop:uniformcase}.

\begin{proof}[Proof of Proposition~\ref{prop:uniformcase}]
Fix $k$ sufficiently large that
all three estimates of
Lemma~\ref{lem:uniform-spectral-estimates}
hold, and abbreviate
\begin{align}
    h=h_k,\qquad g=g_k.
\end{align}

Let $(I_1,\ldots,I_M,K)$ be an
arbitrary admissible Horn tuple,
with index sets of cardinality
$1\leq\ell<N$.

Introduce the averaged incidence weights
\begin{align}
    c_i:=
    \frac1M\sum_{j=1}^M
    \mathbf 1_{\{i\in I_j\}}.
\end{align}
By construction,
\begin{align}
    0\leq c_i\leq1,
    \qquad
    \sum_{i=1}^N c_i=\ell.
\end{align}
In terms of these weights, the
Horn inequality to be established,
Eq.~\eqref{eq:uniform-horn-condition},
reads
\begin{align}
    \sum_{i=1}^N c_i y_i
    \geq\frac{\ell}{N}.
    \label{eq:uniform-weighted-horn}
\end{align}

The admissibility of the tuple
implies the degree identity
\cite{Zelevinsky1999}
\begin{align}
    \sum_{j=1}^M|\lambda(I_j)|
    =|\lambda(K)|.
\end{align}
Recalling that
\begin{align}
    |\lambda(I)|
    =\sum_{i\in I}i
    -\frac{\ell(\ell+1)}2,
\end{align}
and using
$|\lambda(K)|\leq\ell(N-\ell)$,
we obtain
\begin{align}
    E
    &:=
    \sum_{i=1}^N i c_i
    -\frac{\ell(\ell+1)}2
    \nonumber\\
    &=
    \frac1M\sum_{j=1}^M|\lambda(I_j)|
    \nonumber\\
    &=
    \frac{|\lambda(K)|}{M}
    \leq\frac{\ell(N-\ell)}M.
    \label{eq:uniform-index-budget}
\end{align}

The proof below relies only on
the bounds on the weights $c_i$
and the index budget
\eqref{eq:uniform-index-budget}.
No further Littlewood--Richardson
conditions are needed.

Using $\sum_i c_i=\ell$,
we can equivalently write
\begin{align}
\begin{split}
    E={}&
    \sum_{i\leq\ell}
    (\ell-i)(1-c_i)\\
    &+
    \sum_{i>\ell}(i-\ell)c_i.
\end{split}
    \label{eq:uniform-index-identity}
\end{align}
Both sums contain only
nonnegative terms.

We prove
Eq.~\eqref{eq:uniform-weighted-horn}
by distinguishing three ranges
of $\ell$.

\emph{Case 1: $\ell\leq h/2$.}

Since $y_i\geq2/N$ for $i\leq h$,
\begin{align}
    \sum_i c_i y_i
    &\geq\sum_{i\leq h}c_i y_i
    \nonumber\\
    &\geq\frac2N
    \left(\ell-\sum_{i>h}c_i\right).
    \label{eq:uniform-first-range}
\end{align}
From Eq.~\eqref{eq:uniform-index-identity},
\begin{align}
    E\geq(h-\ell)\sum_{i>h}c_i.
\end{align}
Consequently,
\begin{align}
    \sum_{i>h}c_i
    &\leq
    \frac{\ell(N-\ell)}{M(h-\ell)}
    \nonumber\\
    &\leq\frac{2\ell N}{Mh}
    \leq\frac{\ell}{2},
\end{align}
where the last inequality uses
Eq.~\eqref{eq:uniform-condition-h}.

Substitution into
Eq.~\eqref{eq:uniform-first-range}
gives
\begin{align}
    \sum_i c_i y_i
    \geq\frac2N
    \left(\ell-\frac{\ell}{2}\right)
    =\frac{\ell}{N}.
\end{align}

\emph{Case 2: $h/2<\ell<2g$.}

Set $t=\lfloor\ell/2\rfloor$.
Retaining the first $t$ terms,
we obtain
\begin{align}
    \sum_i c_i y_i
    &\geq\sum_{i\leq t}c_i y_i
    \nonumber\\
    &=S_k(t)-
    \sum_{i\leq t}(1-c_i)y_i
    \nonumber\\
    &\geq S_k(t)
    -y_1\sum_{i\leq t}(1-c_i).
    \label{eq:uniform-middle-prefix}
\end{align}

Equation~\eqref{eq:uniform-index-identity}
implies
\begin{align}
    E\geq(\ell-t)
    \sum_{i\leq t}(1-c_i).
\end{align}
Since $\ell-t\geq\ell/2$,
\begin{align}
    \sum_{i\leq t}(1-c_i)
    \leq
    \frac{\ell(N-\ell)}{M(\ell-t)}
    \leq\frac{2N}{M}.
\end{align}

As $t\geq\lfloor h/4\rfloor$,
Eq.~\eqref{eq:uniform-middle-prefix}
and
Eq.~\eqref{eq:uniform-condition-mass}
give
\begin{align}
    \sum_i c_i y_i
    &\geq
    S_k(t)-\frac{2Ny_1}{M}
    \nonumber\\
    &\geq
    S_k\left(\left\lfloor
    \frac h4\right\rfloor\right)
    -2\delta_k
    \nonumber\\
    &\geq\frac{2g}{N}
    >\frac{\ell}{N}.
\end{align}

\emph{Case 3: $\ell\geq2g$.}

We estimate the complement of
the weighted spectral sum:
\begin{align}
    \sum_i c_i y_i
    &=1-\sum_{i\leq g}(1-c_i)y_i
    \nonumber\\
    &\quad-
    \sum_{i>g}(1-c_i)y_i
    \nonumber\\
    &\geq
    1-y_1\sum_{i\leq g}(1-c_i)
    -\frac{N-\ell}{2N},
    \label{eq:uniform-last-range}
\end{align}
where we used
$y_i<1/(2N)$ for $i>g$ and
$\sum_i(1-c_i)=N-\ell$.

From Eq.~\eqref{eq:uniform-index-identity},
\begin{align}
    E\geq(\ell-g)
    \sum_{i\leq g}(1-c_i).
\end{align}
Therefore,
\begin{align}
    \sum_{i\leq g}(1-c_i)
    \leq
    \frac{\ell(N-\ell)}{M(\ell-g)}
    \leq\frac{2(N-\ell)}M.
\end{align}

Substituting this bound into
Eq.~\eqref{eq:uniform-last-range}
and using
Eq.~\eqref{eq:uniform-condition-delta},
we obtain
\begin{align}
    \sum_i c_i y_i
    &\geq
    1-\frac{2y_1(N-\ell)}M
    -\frac{N-\ell}{2N}
    \nonumber\\
    &=
    1-\frac{N-\ell}{N}
    \left(2\delta_k+\frac12\right)
    \nonumber\\
    &\geq\frac{\ell}{N}.
\end{align}

Thus
Eq.~\eqref{eq:uniform-weighted-horn}
holds in all three cases and hence
for every admissible Horn tuple.

By Theorem~\ref{fact:horn},
there exist Hermitian matrices
with spectrum $y/M$ whose sum
has spectrum $u_N$.
Since a Hermitian matrix with
uniform spectrum equals $I_N/N$,
their sum is precisely $I_N/N$.

Each summand is unitarily equivalent
to $\operatorname{diag}(y)/M$.
Absorbing the permutation that
sorts $q^{\otimes k}$, we obtain
\begin{align}
    \frac{I_{d^k}}{d^k}
    =
    \frac1{m^k}\sum_{j=1}^{m^k}
    U_jQ^{\otimes k}U_j^\dagger,
\end{align}
which proves
Proposition~\ref{prop:uniformcase}.
\end{proof}

\section{Uniform probability-tail estimates}
\label{app:tails}

In this appendix, we prove the uniform tail comparison
used in Sec.~\ref{subsec:tensor-power-allocation}.

\subsection{Proof of the uniform tail comparison}
\label{app:uniform-tail-proof}

We prove
Lemma~\ref{lem:uniform-tail-comparison}
using the large-deviation argument
of Aubrun and Nechita
\cite[proof of Proposition~2.5]{AubrunNechita2009}.

Their result establishes eventual
stochastic domination.
We show that the same argument yields
the stronger fixed-factor estimates
required in our setting.

\begin{proof}[Proof of
Lemma~\ref{lem:uniform-tail-comparison}]
We first consider the upper-tail inequality
in Eq.~\eqref{eq:uniform-upper-tail}.

If $\max A<c$, then
$\Pr(\overline A_k\geq v)=0$
for every $v\geq c$, and the
claim is immediate.

Otherwise, let
\begin{align}
    I=[c,\max A],
\end{align}
and define
\begin{align}
    f_k(v)&:=
    \Pr(\overline A_k\geq v)^{1/k},
    \nonumber\\
    g_k(v)&:=
    \Pr(\overline B_k\geq v)^{1/k},
\end{align}
with $v\in I$.

For $X\in\{A,B\}$, define the
cumulant generating function
\begin{align}
    \Lambda_X(\lambda)
    :=
    \log_2\mathbb E[2^{\lambda X}]
\end{align}
and the upper-tail rate function
\begin{align}
    I_X(v)
    :=
    \sup_{\lambda\geq0}
    \{\lambda v-\Lambda_X(\lambda)\}.
\end{align}

Cram\'er's theorem
\cite[Theorem~2.2.3]{DemboZeitouni2010}
implies that, for every $v<\max X$,
\begin{align}
    \lim_{k\to\infty}
    \frac1k\log_2
    \Pr(\overline X_k\geq v)
    =-I_X(v).
\end{align}
For $v>\mathbb E[X]$, this follows
from the large-deviation upper and
lower bounds and the continuity of
the rate function.
For $v\leq\mathbb E[X]$, the
exponential rate is zero, consistently
with $I_X(v)=0$.

Taking $k$th roots, we obtain,
for every $v\in[c,\max A)$,
\begin{align}
    \lim_{k\to\infty}f_k(v)=2^{-I_A(v)},
    \qquad
    \lim_{k\to\infty}g_k(v)=2^{-I_B(v)}.
\end{align}

Since $\max A<\max B$, the same
convergence for $g_k$ also holds
at $v=\max A$.

For $f_k$, the endpoint is treated
directly:
\begin{align}
    f_k(\max A)
    =\Pr(A=\max A)
    =2^{-I_A(\max A)}.
\end{align}
Here the last equality follows from
\begin{align}
    I_A(\max A)
    =-\log_2\Pr(A=\max A).
\end{align}

Consequently, both sequences converge
pointwise on the entire interval $I$ to
\begin{align}
    f(v)&:=2^{-I_A(v)},\\
    g(v)&:=2^{-I_B(v)}.
\end{align}

Both limiting functions are continuous
on $I$. Moreover, $g$ is strictly
positive on $I$, since $\max A<\max B$.

We now show that
\begin{align}
    f(v)<g(v),
    \qquad v\in I,
\end{align}
or, equivalently, that
$I_A(v)>I_B(v)$.

If $v\leq\mathbb E[B]$, then
$I_B(v)=0$, whereas $I_A(v)>0$
since $v\geq c>\mathbb E[A]$.

If $v>\mathbb E[B]$, the inequality
$v\leq\max A<\max B$ ensures that
the supremum defining $I_B(v)$
is attained at some finite
$\lambda_v>0$.
By the strict moment inequality
\eqref{eq:tail-moment-positive},
\begin{align}
    \Lambda_A(\lambda_v)
    <
    \Lambda_B(\lambda_v).
\end{align}
Consequently,
\begin{align}
    I_A(v)
    &\geq
    \lambda_vv-\Lambda_A(\lambda_v)
    \nonumber\\
    &>
    \lambda_vv-\Lambda_B(\lambda_v)
    \nonumber\\
    &=I_B(v).
\end{align}
Thus $f(v)<g(v)$ for every $v\in I$.

By P\'olya's theorem
\cite[Sec.~24, Exercise~24.J]{Bartle1976},
the pointwise convergence of the
nonincreasing functions $f_k$ and $g_k$
to continuous limits on the compact
interval $I$ is uniform.

Since $g$ is continuous and strictly
positive on $I$, the continuity of
inversion in the uniform norm implies
\begin{equation}
    \frac{f_k}{g_k}
    \longrightarrow
    \frac{f}{g}
    \qquad\text{uniformly on }I.
\end{equation}

Continuity and compactness also give
\begin{align}
    r_*:=
    \max_{v\in I}
    \frac{f(v)}{g(v)}
    <1.
\end{align}
Choose $r_*<\theta<1$.
Uniform convergence provides
an integer $k_0$ such that
\begin{align}
    f_k(v)\leq\theta g_k(v),
    \qquad
    v\in I,\quad k\geq k_0.
\end{align}

Raising both sides to the
$k$th power gives
\begin{align}
    \Pr(\overline A_k\geq v)
    \leq
    \theta^k
    \Pr(\overline B_k\geq v).
\end{align}
Increasing $k_0$ if necessary,
we may assume that
$\theta^k\leq1/4$.
For $v>\max A$, the left-hand
side vanishes. We therefore obtain
\begin{align}
    \Pr(\overline A_k\geq v)
    \leq
    \frac14\Pr(\overline B_k\geq v),
    \qquad v\geq c.
\end{align}

The lower-tail inequality follows
by applying the same argument
to the pair $(-B,-A)$, with
separator $-c$.
Indeed, the required strict
moment and endpoint conditions
follow from the corresponding
negative-moment and minimum
conditions for $(A,B)$.

Thus, for all sufficiently large $k$,
\begin{align}
    \Pr(\overline B_k\leq v)
    \leq
    \frac14\Pr(\overline A_k\leq v),
    \qquad v\leq c.
\end{align}

Taking the larger of the two
copy-number thresholds proves
both inequalities simultaneously,
uniformly in $v$.
\end{proof}

\section{Technical lemmas for exact amortization}

\label{app:amortization}

We prove the two technical lemmas used
in Proposition~\ref{prop:amortization}.
The first establishes the absorption
of a fixed auxiliary spectrum by
sufficiently many tensor copies.
The second shows that a uniform
auxiliary state can be canceled
without increasing the one-way
communication cost.

\subsection{Absorption of a fixed auxiliary spectrum}
\label{app:fixed-offset}

\begin{proof}[Proof of Lemma~\ref{lem:offset}]
Let $d_0$ be the common length of $a,b$, and let $h$ be the length of $r$.

We use the following consequence of the large-sample relative-majorization theorem of Farooq et al.~\cite[Corollary~23]{FarooqFritzHaapasaloTomamichel2024}. Let $v,w$ be strictly positive probability vectors of the same length $D$. If
\begin{align}
    D_\alpha(v\Vert u_D)
    &<D_\alpha(w\Vert u_D), \nonumber\\
    D_\alpha(u_D\Vert v)
    &<D_\alpha(u_D\Vert w)
    \label{eq:offset-renyi-criterion}
\end{align}
for every $\alpha\in[1/2,\infty]$, then
\begin{align}
    v^{\otimes s}\prec w^{\otimes s}
\end{align}
for all sufficiently large integers $s$.

Here $D_\alpha$ denotes the Rényi divergence,
\begin{align}
    D_\alpha(v\Vert w)
    :=
    \frac{1}{\alpha-1}
    \log_2\sum_i v_i^\alpha w_i^{1-\alpha},
\end{align}
with its continuous extensions at $\alpha=1,\infty$.

We aim to apply this criterion to
\begin{align}
    v=r\otimes a^{\otimes\ell},
    \qquad
    w=u_h\otimes b^{\otimes\ell},
\end{align}
where $\ell\geq1$ will be chosen so that both inequalities in Eq.~\eqref{eq:offset-renyi-criterion} hold.

For $\alpha\in[1/2,\infty]$, define
\begin{align}
    g_1(\alpha)
    &:=
    D_\alpha(b\Vert u_{d_0})
    -D_\alpha(a\Vert u_{d_0}), \nonumber\\
    g_0(\alpha)
    &:=
    D_\alpha(u_{d_0}\Vert b)
    -D_\alpha(u_{d_0}\Vert a).
\end{align}
We first show that both functions are strictly positive.

For every finite $\alpha\geq1/2$,
\begin{align}
    g_1(\alpha)
    =H_\alpha(a)-H_\alpha(b)>0,
\end{align}
by strict Schur concavity and $a\prec b$.

For $\alpha\neq1,\infty$, we have
\begin{align}
    g_0(\alpha)
    =
    \frac{1}{\alpha-1}
    \log_2
    \frac{\sum_i b_i^{1-\alpha}}
         {\sum_i a_i^{1-\alpha}}.
\end{align}
Since $t^{1-\alpha}$ is strictly concave for $\alpha<1$ and strictly convex for $\alpha>1$, majorization gives $g_0(\alpha)>0$. At $\alpha=1$, the same conclusion follows from strict concavity of the logarithm:
\begin{align}
    g_0(1)
    =
    \frac{1}{d_0}
    \sum_i\log_2\frac{a_i}{b_i}>0.
\end{align}

Finally, the assumptions on the extreme entries imply
\begin{align}
    g_1(\infty)
    &=\log_2\frac{b_{\max}}{a_{\max}}>0, \nonumber\\
    g_0(\infty)
    &=\log_2\frac{a_{\min}}{b_{\min}}>0.
\end{align}

Both functions are continuous on the compactified interval $[1/2,\infty]$. Therefore,
\begin{align}
    \delta:=
    \min_{\alpha\in[1/2,\infty]}
    \min\{g_0(\alpha),g_1(\alpha)\}>0.
    \label{eq:offset-uniform-margin}
\end{align}

We now choose $\ell$ so that the Rényi gaps accumulated by the tensor powers dominate the fixed contribution of $r$.

Set
\begin{align}
    C:=
    \max\left\{
    D_\infty(r\Vert u_h),
    D_\infty(u_h\Vert r)
    \right\},
\end{align}
and choose an integer $\ell\geq1$ such that $\ell\delta>C$.

Recall that
\begin{align}
    v=r\otimes a^{\otimes\ell},
    \qquad
    w=u_h\otimes b^{\otimes\ell},
\end{align}
with common dimension $D=hd_0^\ell$.

By additivity of Rényi divergences,
\begin{align}
    &D_\alpha(w\Vert u_D)
    -D_\alpha(v\Vert u_D) \nonumber\\
    &\qquad=
    \ell g_1(\alpha)-D_\alpha(r\Vert u_h)
    \nonumber\\
    &\qquad\geq\ell\delta-C>0,
    \label{eq:offset-first-gap}
\end{align}
and
\begin{align}
    &D_\alpha(u_D\Vert w)
    -D_\alpha(u_D\Vert v) \nonumber\\
    &\qquad=
    \ell g_0(\alpha)-D_\alpha(u_h\Vert r)
    \nonumber\\
    &\qquad\geq\ell\delta-C>0.
    \label{eq:offset-second-gap}
\end{align}
Here we used the monotonicity of Rényi divergence in its order.

Thus $v,w$ satisfy both inequalities in Eq.~\eqref{eq:offset-renyi-criterion} for every $\alpha\in[1/2,\infty]$.

By the cited large-sample relative-majorization criterion, for all sufficiently large integers $s$ there exists a stochastic matrix $T_s$ such that
\begin{align}
    T_s w^{\otimes s}=v^{\otimes s},
    \qquad
    T_s u_D^{\otimes s}=u_D^{\otimes s}.
\end{align}
Since $T_s$ preserves the uniform distribution, it is doubly stochastic. Hence
\begin{align}
    v^{\otimes s}\prec w^{\otimes s}.
\end{align}

Choosing such an $s$ and setting $K=\ell s$, we obtain
\begin{align}
    r^{\otimes s}\otimes a^{\otimes K}
    \prec
    u_h^{\otimes s}\otimes b^{\otimes K}.
    \label{eq:offset-tensor-majorization}
\end{align}

It remains to reduce the $s$ copies of $r$ to a single copy. Since $u_h\prec r$, tensor-product monotonicity gives
\begin{align}
    r\otimes u_h^{\otimes(s-1)}
    \prec r^{\otimes s}.
\end{align}
Combining this with Eq.~\eqref{eq:offset-tensor-majorization}, we find
\begin{align}
\begin{aligned}
    &(r\otimes a^{\otimes K})
    \otimes u_h^{\otimes(s-1)}\prec
    (u_h\otimes b^{\otimes K})
    \otimes u_h^{\otimes(s-1)}.
\end{aligned}
\end{align}

A common uniform tensor factor can be canceled from ordinary majorization: the partial-sum inequalities at multiples of its dimension recover precisely those of the original vectors.

Consequently,
\begin{align}
    r\otimes a^{\otimes K}
    \prec
    u_h\otimes b^{\otimes K},
\end{align}
as required.
\end{proof}

\subsection{Cancellation of a uniform auxiliary state}
\label{app:uniform-cancel}
\begin{proof}[Proof of Lemma~\ref{lem:uniform-cancel}]
Let $X=\operatorname{diag}(x)$ and
$Y=\operatorname{diag}(y)$.

Tensoring any unitary-orbit decomposition
with $I_h/h$ immediately gives
\begin{align}
    \Bto(x\otimes u_h\to y\otimes u_h)
    \leq\Bto(x\to y).
\end{align}

To establish the reverse inequality,
suppose that the transformation
$x\otimes u_h\to y\otimes u_h$
admits a decomposition with $m$ terms:
\begin{align}
    X\otimes\frac{I_h}{h}
    =
    \sum_{j=1}^{m}w_jU_j
    \left(Y\otimes\frac{I_h}{h}\right)
    U_j^\dagger,
    \label{eq:uniform-orbit-decomposition}
\end{align}
where $w_j\geq0$ and $\sum_jw_j=1$.

Our goal is to construct unitaries
$V_1,\ldots,V_m$ such that
\begin{align}
    X=\sum_{j=1}^{m}
    w_jV_jYV_j^\dagger.
    \label{eq:uniform-canceled-decomposition}
\end{align}
This would give a protocol
with the same message alphabet size
and the same weights.

Multiplying
Eq.~\eqref{eq:uniform-orbit-decomposition}
by $h$, we obtain an additive
eigenvalue problem. Its summands
have spectra
\begin{align}
    \operatorname{rep}_h(w_1y),
    \ldots,
    \operatorname{rep}_h(w_my),
\end{align}
and their sum has spectrum
$\operatorname{rep}_h(x)$.

Here $\operatorname{rep}_h(a)$
denotes the list obtained by repeating
each entry of $a$ exactly $h$ times.

We therefore need to establish that
repeating every eigenvalue $h$ times
does not change the feasibility of
an additive spectral decomposition.

\emph{Cancellation of repeated spectra.}

Let $a^{(1)},\ldots,a^{(m)},c$
be decreasing real lists of length $D$.
Suppose that Hermitian matrices with
spectra
\begin{align}
    \operatorname{rep}_h(a^{(1)}),
    \ldots,
    \operatorname{rep}_h(a^{(m)})
\end{align}
can have a sum with spectrum
$\operatorname{rep}_h(c)$.

We show that Hermitian matrices
with spectra $a^{(1)},\ldots,a^{(m)}$
can have a sum with spectrum $c$.

By the additive Horn theorem,
Theorem~\ref{fact:horn}, it suffices
to verify the trace condition
and every admissible Horn inequality
for the unrepeated spectra.

Consider an arbitrary admissible Horn tuple $(I_1,\ldots,I_m,K)$ in dimension $D$, with all index sets of cardinality $1\leq r<D$. Our goal is to recover its Horn inequality from the feasibility of the repeated spectra.

To this end, for every index set $I\subseteq\{1,\ldots,D\}$, define
\begin{align}
    I^{[h]}:=
    \bigcup_{i\in I}
    \{h(i-1)+1,\ldots,hi\}.
    \label{eq:horn-index-lift}
\end{align}
The lifted set has cardinality $hr$ and belongs to $\{1,\ldots,hD\}$.

We first establish that the lifted tuple $(I_1^{[h]},\ldots,I_m^{[h]},K^{[h]})$ is admissible in dimension $hD$.

Recall that for $I=\{i_1<\cdots<i_r\}$, the associated partition is
\begin{align}
    \lambda(I)=(i_r-r,\ldots,i_1-1).
\end{align}
Writing $I^{[h]}=\{j_1<\cdots<j_{hr}\}$, we have
\begin{align}
    j_{h(a-1)+s}=h(i_a-1)+s,
    \qquad 1\leq a\leq r,\quad 1\leq s\leq h.
\end{align}
Consequently,
\begin{align}
    j_{h(a-1)+s}-[h(a-1)+s]
    =h(i_a-a),
\end{align}
which proves
\begin{align}
    \lambda(I^{[h]})
    =\operatorname{rep}_h\bigl(h\lambda(I)\bigr).
    \label{eq:horn-partition-lift}
\end{align}

It remains to show that this transformation preserves the positivity of the Littlewood--Richardson coefficients defining admissibility.

For a partition $\lambda$, let $\lambda'$ denote its conjugate. The transformation in Eq.~\eqref{eq:horn-partition-lift} can be expressed as
\begin{align}
    \operatorname{rep}_h(h\lambda)
    =
    \bigl(h(h\lambda)'\bigr)'.
\end{align}

We use two standard properties of Littlewood--Richardson coefficients: invariance under simultaneous conjugation of all partitions~\cite[Ch.~I]{Macdonald1995}, and preservation of positivity under simultaneous multiplication by a positive integer, which follows from the semigroup property~\cite[Theorem~1]{Zelevinsky1999}.

Starting from
\begin{align}
    c_{\lambda(I_1),\ldots,\lambda(I_m)}^{\lambda(K)}>0,
\end{align}
we may therefore successively scale all partitions by $h$, conjugate them, scale again by $h$, and conjugate once more. Using Eq.~\eqref{eq:horn-partition-lift}, this yields
\begin{align}
    c_{\lambda(I_1^{[h]}),\ldots,\lambda(I_m^{[h]})}^{\lambda(K^{[h]})}>0.
\end{align}
For multiple factors, preservation under scaling follows by applying the two-factor result to a chain of positive Littlewood--Richardson coefficients.

Moreover, the lifted partitions fit inside the $hr\times h(D-r)$ rectangle corresponding to dimension $hD$.

Thus the lifted tuple is admissible, as claimed.

Applying its Horn inequality
to the repeated spectra gives
\begin{align}
\begin{aligned}
    &\sum_{i\in K^{[h]}}
    \operatorname{rep}_h(c)_i\leq
    \sum_{j=1}^{m}
    \sum_{i\in I_j^{[h]}}
    \operatorname{rep}_h(a^{(j)})_i.
\end{aligned}
\end{align}

By construction, each selected
eigenvalue appears exactly $h$
times. Therefore,
\begin{align}
    h\sum_{i\in K}c_i
    \leq
    h\sum_{j=1}^{m}
    \sum_{i\in I_j}a_i^{(j)}.
\end{align}
Dividing by $h$, we obtain
\begin{align}
    \sum_{i\in K}c_i
    \leq
    \sum_{j=1}^{m}
    \sum_{i\in I_j}a_i^{(j)},
\end{align}
which is precisely the original
Horn inequality.

The trace condition is recovered
in the same way. Since the argument
applies to every admissible tuple,
Theorem~\ref{fact:horn} proves the
claimed spectral cancellation.

\emph{Conclusion.}

We can now return to
Eq.~\eqref{eq:uniform-orbit-decomposition}.

Applying the spectral cancellation
just established with
\begin{align}
    a^{(j)}=w_jy,
    \qquad c=x,
\end{align}
we obtain Hermitian matrices
$H_1,\ldots,H_m$ satisfying
\begin{align}
    \spec(H_j)=w_jy,
    \qquad
    \spec\left(\sum_jH_j\right)=x.
\end{align}

After a common unitary conjugation,
we may assume
\begin{align}
    \sum_jH_j=X.
\end{align}
Since each $H_j$ has spectrum
$w_jy$, there exists a unitary
$V_j$ such that
\begin{align}
    H_j=w_jV_jYV_j^\dagger.
\end{align}

Consequently,
\begin{align}
    X=\sum_{j=1}^{m}
    w_jV_jYV_j^\dagger,
\end{align}
which is exactly the decomposition
we sought.

By the operational characterization
in Appendix~\ref{app:operational},
this gives
\begin{align}
    \Bto(x\to y)
    \leq
    \Bto(x\otimes u_h\to y\otimes u_h).
\end{align}

Together with the opposite inequality,
this proves
\begin{align}
    \Bto(x\otimes u_h\to y\otimes u_h)
    =
    \Bto(x\to y).
\end{align}
\end{proof}

\ifCLASSOPTIONcaptionsoff
  \newpage
\fi

\bibliographystyle{IEEEtran-linked}
\bibliography{IEEEabrv,references-ieee-linked}
\end{document}